\documentclass{aa}
\usepackage[dvips]{graphics}
\usepackage[dvips]{rotating}
\begin{document}

   \thesaurus{03 (13.25.2; 11.19.1; 11.01.2)}
   \title{Heavy obscuration in X--ray weak AGNs}

   \subtitle{}

   \author{R. Maiolino \inst{1} \and
          M. Salvati \inst{1} \and
	  L. Bassani \inst{2} \and
	  M. Dadina  \inst{2} \and
	  R. Della Ceca \inst{3} \and
	  G. Matt \inst{4} \and
	  G. Risaliti \inst{5} \and
	  G. Zamorani \inst{6}
          }

   \offprints{R. Maiolino}

   \institute{Osservatorio Astrofisico di Arcetri, Largo E.Fermi 5,
              I-50125, Firenze, Italy, maiolino@arcetri.astro.it,
              salvati@arcetri.astro.it
         \and
             Istituto Tecnologie e Studio Radiazioni Extraterrestri, CNR,
	     Via Gobetti 101, I-40129 Bologna, Italy,
         loredana@botes1.tesre.bo.cnr.it, dadina@terra.tesre.bo.cnr.it
	 \and
	     Osservatorio Astronomico di Brera, Via Brera 28, I-20121 Milano,
		Italy, rdc@brera.mi.astro.it
	 \and
	     Dipartimento di Fisica, Universit\`a di Roma III, Via della
		Vasca Navale 84, I-00146, Roma, Italy, matt@haendel.fis.uniroma3.it
	\and
		Dipartimento di Astronomia, Universit\`a di Firenze, Largo E.Fermi 5,
			I-50125, Firenze, Italy, risaliti@arcetri.astro.it
	\and
	     Osservatorio Astronomico di Bologna, Via Zamboni 33,
		I-40126 Bologna, Italy, zamorani@astbo3.bo.astro.it
             }

   \date{Received 13 March 1998 / Accepted }

   \maketitle

   \begin{abstract}

We present observations in the 0.1--100 keV spectral band of 8 
Seyfert 2 galaxies, obtained by means of BeppoSAX. These
sources were selected according to their [OIII] optical emission line flux,
to avoid biases against obscuration on the pc scale.
All sources were detected. All of them are weak X--ray emitters,
and most of them are characterized by
prominent iron lines at 6.4--7 keV (EW$> 500$ eV)
and by a flat continuum, indicative of heavy
obscuration along our line of sight (N$_H > 10^{25}$ cm$^{-2}$ in most cases).
These results 1) provide further evidence in favor of the
unified scenario, and 2) indicate that
the average obscuration of type 2 AGNs is very likely much higher than
deduced by former X--ray surveys. These findings have important
implications for the synthesis of the X--ray background.

   \end{abstract}

%

\section{Introduction} \label{intro}

The spectra of active galactic nuclei (AGNs) at high energies ($>$3 keV)
provide important information on the physics of the nuclear region.
Unfortunately, the limited sensitivity of current and past X--ray missions
has usually restricted the investigation to bright AGNs. Little is known
about the hard X--ray properties of low luminosity AGNs. This problem
concerns particularly Seyfert 2s,
since they are generally weaker than
their type 1 counterparts. Yet, the hard X--ray properties of type 2 Seyferts
are most interesting, since they provide important information on the 
obscuration affecting this class of objects.

Several observational data indicate that type 2 Seyfert nuclei suffer
significant obscuration along our line of sight. The nature of the
obscuring medium is matter of debate. The unified model (Antonucci
\cite{antonucci})
ascribes the obscuration of Sy2 nuclei to a gaseous pc-scale circumnuclear
torus. According to this model Sy1s and Sy2s would be identical physical
objects, while the orientation of the line of sight with respect to the torus 
would be responsible for the obscuration of the BLR and of the
nuclear engine (X--UV source) in type 2 Seyferts.
Hard X--ray spectra are probably the best observational tool to directly
measure the absorption affecting Sy2 nuclei. Indeed, Sy2 spectra in the
2--10 keV range show evidence for a power law component similar to that
observed in Sy1s (photon index $\sim$ 1.7) and a cutoff due to photoelectric
absorption. The latter indicates absorbing column densities between $10^{22}$
and $10^{24}$cm$^{-2}$ (e.g. Awaki \cite{awaki2}, Ueno et al. 
\cite{ueno2}, Smith \&
Done \cite{smith}, Turner et al. \cite{turner_a}),
often ascribed to the obscuring
torus.

In some sources the absorbing column density is so high (N$_H>10^{24}$cm$^{-2}$)
that it is optically thick to Compton scattering. In this case the direct
component is completely absorbed in the 2--10 keV range.
However, the nuclear radiation can be
Compton reflected by cold material surrounding the X--ray source
(possibly the same torus responsible for the obscuration) or
scattered by free electrons in a highly ionized
warm gas.  If these media extend outside the
absorbing torus, then signatures of the X--ray nuclear activity are
observable via the reflected/scattered continuum. Also, the cold reflecting
medium produces a fluorescence iron line at 6.4 keV,
while He-- and H--like iron in the warm 
scattering medium should emit lines at 6.7 and 6.96 keV. Since the
direct continuum is completely suppressed in Compton thick sources, these
iron lines are characterized by equivalent widths ($\geq 1$ keV)
larger than in Sy1s or in Compton thin Sy2s. Until a few years ago NGC1068
was the only Compton thick source known. A few more sources of this class
have been discovered recently (see Matt \cite{matt_b} for a review).

As anticipated above, Sy2s in
early spectroscopic studies were mostly selected from former 
all--sky X--ray surveys, hence these studies were generally biased for
X--ray bright objects.
Very likely, this selection criterion resulted in a bias
in favor of low N$_H$ Sy2s.
Later on, hard X--ray spectroscopic studies (mostly
by means of ASCA) probed fainter samples of AGNs (e.g. Turner et al. 
\cite{turner_a}).
However, many of the Sy2s observed by ASCA were selected amongst sources
known to show broad lines in polarized light (Awaki et al.
\cite{awaki}). This selection
criterion might introduce a bias for low N$_H$ as well. Heisler et al.
(\cite{heisler})
showed that the detectability of polarized broad lines is related to
the obscuration of the nuclear region.

Summarizing, former X--ray spectroscopic surveys were seriously biased
against heavily obscured Sy2s and, therefore, they are not suitable to study
the real distribution of the absorbing column densities N$_H$.

The knowledge of the distribution of N$_H$ in Sy2s is important to
understand the nature of their obscuring medium, which has implications
for the unified model. Also, the distribution of N$_H$ is relevant to the
synthesis of the X--ray background. Indeed, obscured AGNs are thought to
contribute to most of the high energy ($>$ 2 keV) extragalactic
background (Comastri et al. \cite{comastri},
Madau et al. \cite{madau}).

\section{The weak Sy2 BeppoSAX core program} \label{prog_des}

We have undertaken a program of observations with BeppoSAX, the Italian--Dutch
X--ray satellite, aimed at studying the hard X--ray properties of weak Sy2s
and at assessing the ``real'' distribution of their
absorbing column densities.

As described in the next section, BeppoSAX is an excellent tool to pursue this
goal, since it combines high sensitivity (required to observe weak AGNs)
and a wide spectral coverage (0.1--300 keV, required to identify and disentangle
various spectral components).

Sy2s suitable for this study were drawn out of Maiolino \& Rieke's
(\cite{maiolino_a}) sample.
This sample is extracted from the Revised Shapley-Ames (RSA) catalog of galaxies
(that is limited to magnitude $B_T < 13.2$, Sandage \& Tammann \cite{sandage}),
and Seyfert galaxies are selected
according to their optical lines. As discussed in Maiolino \& Rieke, this
sample is much less biased than others, both in terms of luminosity of the
Seyfert nuclei and in terms of properties of their host galaxies.
Out of the 54 Sy2s in the Maiolino \& Rieke sample 22 have already been
observed by ASCA. We selected 8 of the remaining 32 Sy2s, based on their
[OIII]5007\AA
 (narrow) line flux: to maximize the chances of detection we chose the
sources showing the highest [OIII] flux.
 Maiolino \& Rieke also show that, although
the [OIII] line is emitted on scales much larger than the putative pc-scale
torus, the [OIII] is not a completely isotropic indicator of the nuclear
luminosity, for the host galaxy disk might obscure part of the NLR.
However, once the [OIII] flux is corrected for the extinction deduced from
the Balmer decrement, it should provide an indication of the nuclear activity
that is independent of the pc-scale obscuration due to the
torus. As a consequence, although our [OIII]-based selection criterion
might introduce a bias in our sample for {\it intrinsically}
luminous sources, it
avoids biases against highly obscured Sy2 nuclei, thus overcoming
limitations of former surveys.

Preliminary results of this survey were published in Salvati et al.
(\cite{salvati}, \cite{salvati2}).
In this paper we report and discuss results of BeppoSAX observations for all
of the 8 Seyfert 2s in our [OIII] selected sample. A more thorough
statistical analysis, obtained by merging our BeppoSAX data with data in
the literature, is presented in Bassani et al. (in prep.).

\section{Observations and data reduction} \label{obs}

A description of the BeppoSAX observatory is given in Boella et al.
(\cite{boella_a}).
The payload instruments include four
co--aligned narrow field instruments: a Low Energy Concentrator
Spectrometer (LECS, Parmar et al. \cite{parmar}),
three Medium Energy Concentrator 
Spectrometers (MECS, Boella et al. \cite{boella_b}),
 a High Pressure Gas Scintillation Proportional
Counter (HPGSPC, Manzo et al. \cite{manzo})
and a Phoswich Detector System (PDS, Frontera et al. \cite{frontera}).
Both LECS and MECS spectrometers
have imaging
capabilities (angular resolution $\sim$ 1.2 arcmin, FWHM) and cover the
0.1--10 keV and 1.5--10 keV spectral bands, respectively (in the overlapping
spectral region the MECS are three times more sensitive than LECS). Their
energy resolution is about 8\% at 6 keV. HPGSPC and PDS operate in the
4--120 keV and 15--300 keV spectral bands, respectively. In the overlapping
region the PDS is more sensitive (by a factor of 4--8), while the HPGSPC has
superior energy resolution.

Table \ref{tab_obs} lists the sources observed in our program so far, along
with the on--source total integration time and net count rate (i.e.
background subtracted) for each
instrument.
One of the MECS units stopped working in May 1997, as a consequence
MCG-05-18-002 was observed with two MECS units only.

\begin{table*}
\caption[]{Observations log}\label{tab_obs}
\begin{tabular}{lccccccc}
\hline
\hline
Source & Observation start & \multicolumn{2}{c}{LECS (0.1-4keV)} &
\multicolumn{2}{c}{MECS (1.65-10.5 keV)} &
\multicolumn{2}{c}{PDS (15-100 keV)}\\
 & Date &  Duration$^a$  & net counts$^b$ &
	       Duration$^a$  & net counts$^b$ &
	       Duration$^a$  & net counts$^b$ \\
\hline
NGC1386 & 10/12/96 &  10.3 & 51$\pm$8.6 & 30.2 & 158$\pm$42 & --$^c$ &
  --$^c$ \\
NGC2273 & 22/02/97 & 10.0 & 30$\pm$9 & 24.2 & 318$\pm$23 & 12 & $<$1150\\
NGC3081 & 20/12/96 & 8.9 & 52$\pm$12 & 15.4 & 241$\pm$20& 7 & $<$870 \\
NGC3393 & 08/01/97 & 6.4 & 6$\pm$10 & 14.7 & 81$\pm$15& 6 & 1509$\pm$492 \\
NGC4939 & 27/01/97 & 13.6 & 128$\pm$16 & 33.3 & 608$\pm$30 & 15.8 &
  3542$\pm$823 \\
NGC4941 & 22/01/97 & 14.3 & 48$\pm$14 & 28.8 & 245$\pm$23 & 14 & 1325$\pm$642\\
NGC5643 & 01/03/97 & 7.4 & 52$\pm$8 & 10.4 & 237$\pm$18 & 5 & $<$430 \\
MCG-05-18-002 & 21/11/97 & 3.6 & 16$\pm$7 & 8.9$^d$ &
  60$\pm$11$^d$ & 4 & $<$ 690 \\
\hline
\end{tabular}

The uncertainties on the fluxes are 1$\sigma$.
Notes:\vspace{-0.1truecm}
\begin{list}{}{}
\item $^a$ In ksec.
\item $^b$ (Source -- background) $\pm~1~\sigma$.
\item $^c$ During the observation of NGC 1386 the PDS beam (1.4$^{\circ}$
wide) also included NGC 1365; the detected flux is dominated by the
latter (Maiolino et al. in preparation).
\item $^d$ This observation was performed with two MECS units only.
\end{list}
\end{table*}

For all sources, but NGC 5643,
LECS and MECS spectra were extracted from an aperture of 8$'$ and 4$'$ in
radius, since these apertures
 were found to optimize the signal--to--noise ratios for
these faint sources. For NGC 5643 we chose an extraction radius of 1$'$,
since outside of this radius the observed flux is affected by the emission
of a nearby galaxy cluster (see Sect.~\ref{results}).
Another exception is NGC 1386, where
we used for the LECS the same extraction aperture
used for the MECS (i.e. 4$'$ in
radius) to avoid different contributions to the soft X--ray flux
from the surrounding Fornax cluster.
The background was extracted by using the same apertures in blank sky
observation files.

The PDS consists of four phoswich units, two pointing at the
source and the other two 210$'$ away. The two pairs switch on and off
source every 96 seconds. The net count rate is obtained by subtracting
the ``off'' counts from the ``on'' counts.

We rebinned the spectra to have a minimum of 10 counts per channel
(source $+$ background), so that
statistics can be treated in the Poissonian limit and the energy resolution
is not affected significantly.

We did not detect short term variability in any of our sources.
However, the signal--to--noise is too low to provide stringent constraints
since these sources are weak and the integration times relatively short.

\section{Spectral analysis and constraints on N$_H$} \label{analysis}

\subsection{Continuum fitting} \label{cont}

We used three models to fit the observed continuum.

\begin{enumerate}

\item Transmission model (Compton thin). It consists of a power law (similar
to that observed in Sy1s) transmitted through an absorbing cold medium.

\item Warm scattering model (Compton thick). This model assumes that the primary
power law is completely absorbed along our line of sight by a medium
that is  thick to Compton scattering. However, the primary radiation is
scattered into our line of sight by a warm, highly ionized
 gas located outside the absorbing medium. As
a consequence, this model consists of a power law with no absorption
in excess of the Galactic value.

\item Cold reflection model (Compton thick). As in model 2, here the
primary radiation is completely absorbed along our line of sight by Compton
thick material. However, in this model the primary radiation is Compton
reflected into our line of sight by a cold, neutral material in the
circumnuclear region, possibly identified with the same torus
responsible for the obscuration. In this case, the spectral model consists
of the ``hump'' characteristic of the cold reflection.
In the 2--10 keV range, the continuum is much flatter than the primary
radiation.

\end{enumerate}

\noindent Since the sources in our sample are weak, several of our 
spectra have low signal--to--noise ratio in the continuum.
Therefore, we reduced to the minimum possible the
number of free parameters. In particular,
we froze the photon index $\Gamma$ of the continuum to 1.7 in all
models. This is the average photon index in Sy1
galaxies  and it is thought to arise from a primary power law
with photon index 1.9 flattened, in the 2-20 keV range, by a cold reflection
component due to reprocessing by the accretion disk (Nandra \& Pounds 
\cite{nandra};
Nandra et al. \cite{nandra2}).

However, to allow for a comparison with other studies in the literature where
$\Gamma$ is a free parameter, in the Appendix we also report results of
transmission model fits with this additional degree of freedom.

In most cases the extrapolation of the continuum model
fitted at high energies ($>3$ keV) falls short of accounting for the emission
in the soft X--ray range ($<3$ keV). Therefore, we also introduced
a black body component (kT $\approx$ a few 0.1 keV) to account
for this ``soft excess''. The black body is just an analitical form
that fits nicely the
spectrum below $\sim 3$ keV, but it does not necessary reflect
the real nature of this flux; this will be
discussed in Sect.~\ref{disc_soft}.

However, only data above 3 keV were used to statistically discriminate
between the transmission, warm scattering and cold reflection
models, so that the $\chi ^2$ were little affected by the soft excess.

In all models, Galactic photoelectric absorption is included.

\subsection{The Fe line} \label{fe}

The hard X--ray spectrum of Seyfert galaxies is usually characterized by
a fluorescent line at 6.4 keV produced by low ionization iron (i.e. iron
with no vacancies in the L~shell, and then less ionized than Fe {\sc xvii}).
Also, in some cases (e.g. NGC 1068, Ueno et al. 
\cite{ueno})
recombina\-tion/re\-sonant emission from He--like (6.7 keV) and H--like iron
(6.96 keV) is observed.

Our fit includes an unresolved gaussian, whose central energy
is left free, to account for the iron line.
In some cases there are indications that the iron line is resolved; in these
cases we tried to fit the line with two gaussians centered
at 6.4+6.7 keV (neutral and He--like iron) or at 6.4+6.96 keV (neutral and H--like
iron).
The two-components fit was considered
significant only if the $\chi ^2$ probability
improved at a confidence level of at least 90\% .

The equivalent width (EW) of the fluorescence 6.4 keV Fe K$\alpha$ 
line provides useful
information to constrain the column density that absorbs the continuum and,
in some cases, to distinguish between the
Compton thick and the Compton thin cases.
This fluorescence line is thought to arise from the accretion
disk (Lightman \& White \cite{lightman};
Fabian et al. \cite{fabian}) and, possibly, 
from the torus as well (Ghisellini et al. \cite{ghisellini}), with 
EW(Fe K$\alpha$)$<$ 200--300 eV.
Cold absorbing column densities N$_H<10^{23}$cm$^{-2}$
do not change significantly
the observed
EW(Fe K$\alpha$), since the photoelectric cutoff occurs at energies lower 
than 6 keV.
The presence of an absorbing medium with N$_H>10^{23}$ depresses the
continuum beneath the iron lines and, if the line 
is produced by material more extended than the
obscuring medium, the EW increases. Moreover, 
line photons are produced in the absorber itself, whose EW may be
significant provided that the covering factor is large enough.
EW(Fe K$\alpha$) $ > $ 1 keV is
characteristic of purely reflected/scattered spectra and, therefore, indicates
complete absorption of the primary radiation, 
and then is used to  identify Compton thick candidates.
Interpreting EW(Fe K$\alpha$) $>$ 1keV
with a Compton thin, low absorption model would
require an iron gas abundance several times higher than the cosmic value
(e.g. Matt et al. \cite{matt_c}).

The K$\alpha$ lines from He-- and H--like iron 
(6.7 and 6.96 keV respectively) are emitted from highly ionized gas.
Their EW with respect to the scattered
component can be as high as a few keV (Matt et al. \cite{matt_a}), 
while, if the primary radiation is visible, they are likely to be diluted to
invisibility. Therefore, detection of lines with 
EW(Fe$_{6.7+6.96}$) $ >$ 1 keV is a clue that the primary radiation
is heavily absorbed, i.e. N$_H>10^{24}$cm$^{-2}$.
However, determining the properties of the obscuring
material and of the scattering medium from the EW of these high
ionization lines presents more uncertainties than the 6.4 keV line.
Indeed, the intensity of the He-- and H-- lines
depends non linearly from the optical thickness of the scattering gas.
Also, starburst activity can increase the intensity of these lines.
On the other hand cases where the 6.7--6.96 keV lines dominate are rare:
in most Sy2s the 6--7 keV region is dominated by the
line at 6.4 keV (Turner et al. \cite{turner_a},
also this paper Sect.~\ref{results}).

Finally, we should mention that, both in terms of EW (Fe K$\alpha$) and 
continuum shape, objects obscured by a column density
$10^{24}<$N$_H<10^{25}$cm$^{-2}$ look like
purely reflected sources in the 2--10 keV range
(ASCA); however, their Compton thickness, even if larger than unity, is still
small enough to permit transmission of a significant fraction of photons 
in the 10--100 keV range. So far, only NGC4945 and Mkn3 are
known to belong to this
class (Done et al. \cite{done}, Turner et al. \cite{turner_b}).
Such class of objects would
be readily recognized in the 10--100 keV range of BeppoSAX data (PDS).

\subsection{Other nuclear absorption indicators} \label{other_abs}

Various authors have also used other methods to determine
the nuclear absorption along our line
of sight and, specifically, to distinguish between
Compton thin and Com\-pton thick sources.
By assuming the unified model correct, the ratio between
the hard X--ray luminosity and an isotropic indicator
of the intrinsic luminosity should
provide indications on the amount of absorption affecting the nuclear
X--ray source.
The luminosity of the [OIII] line can be considered an isotropic indicator
of the nuclear intrinsic luminosity, though caveats
discussed in Sect.~\ref{prog_des} must be taken into account
(in particular the [OIII] luminosity must be corrected for the NLR
extinction as deduced from the Balmer decrement). 
The effect of a high absorbing column density is to lower the
L$_{X(2-10keV)}$/L$_{[OIII]}$ ratio with respec to Sy1s. The reduction
is at most by a factor of $\sim~5$ when N$_H <$ a few times $10^{23}$cm$^{-2}$,
and by about two orders of magnitude when N$_H>10^{24}$cm$^{-2}$.

Mulchaey et al. (\cite{mulchaey}) and Alonso-Herrero et al. \cite{herrero}
used the L$_{X(2-10keV)}$/L$_{[OIII]}$
ratio to identify absorption effects in Sy2s.
They do not find significant differences between Sy2s and Sy1s (except for
NGC1068). However, their Sy2
sample is seriously biased toward X--ray bright source
(see Sect.~\ref{intro} and Sect.~\ref{prog_des}),
hence little absorbed;
also, they do not correct L$_{[OIII]}$ for extinction. Turner et al.
(\cite{turner_b})
adopt this method on a better selected sample
(by including weaker sources with respect to
the Mulchaey et al. sample). As a result, they identify some Sy2s
that are suspected to be reflection
dominated, based on a L$_{X(2-10keV)}$/L$_{[OIII]}$ much lower than observed
in Sy1s.
We will use the
L$_{X(2-10keV)}$/L$_{[OIII]}$ ratio (where L$_{[OIII]}$ is corrected for
extinction in the narrow line region)
as an aid to identify heavily absorbed sources. The distribution of
L$_{X(2-10keV)}$/L$_{[OIII]}$ for the sources in our sample, compared
to a sample of Sy1s, is shown in Fig.\ref{fig_xoiii} ([OIII] data are
from Dadina et al. \cite{dadina}).
A more thorough analysis of the 
L$_{X(2-10keV)}$/L$_{[OIII]}$ ratio for a large sample of Sy2s, including
the ones presented in this paper, is discussed in Bassani et al. (in prep.).

Mulchaey \cite{mulchaey},
Mas-Hesse et al. \cite{mas_hesse} and Awaki (\cite{awaki2})
use also the Far--IR (FIR) emission
(as deduced from the 60$\mu m$ and 100$\mu m$
IRAS data) as isotropic indicator of the intrinsic nuclear luminosity,
hence the
L$_{X(2-10keV)}$/L$_{FIR}$ as indicator of nuclear absorption.
However, in these low luminosity
AGNs the FIR luminosity is often dominated
by star formation in the host galaxy (Maiolino et al. \cite{maiolino_b}).
Therefore, we do not consider the FIR emission a reliable
indicator of the AGN luminosity.

Rapid variability (on scales of a few 10 ksec)
of the X--ray continuum is indicative that the
observed radiation is primary emission seen directly and not 
reprocessed by pc--scale reflecting media, i.e. that
the source is Compton thin.
Unfortunately, as discussed in Sect.~\ref{obs}, our data do not provide good
constraints on the short term variability. Yet, the lack of significant
long term
variability (i.e. on a time scale of a few years) is in favor of a
reprocessed origin of the observed emission.

\begin{figure*}[!]
\resizebox{\hsize}{!}{\includegraphics{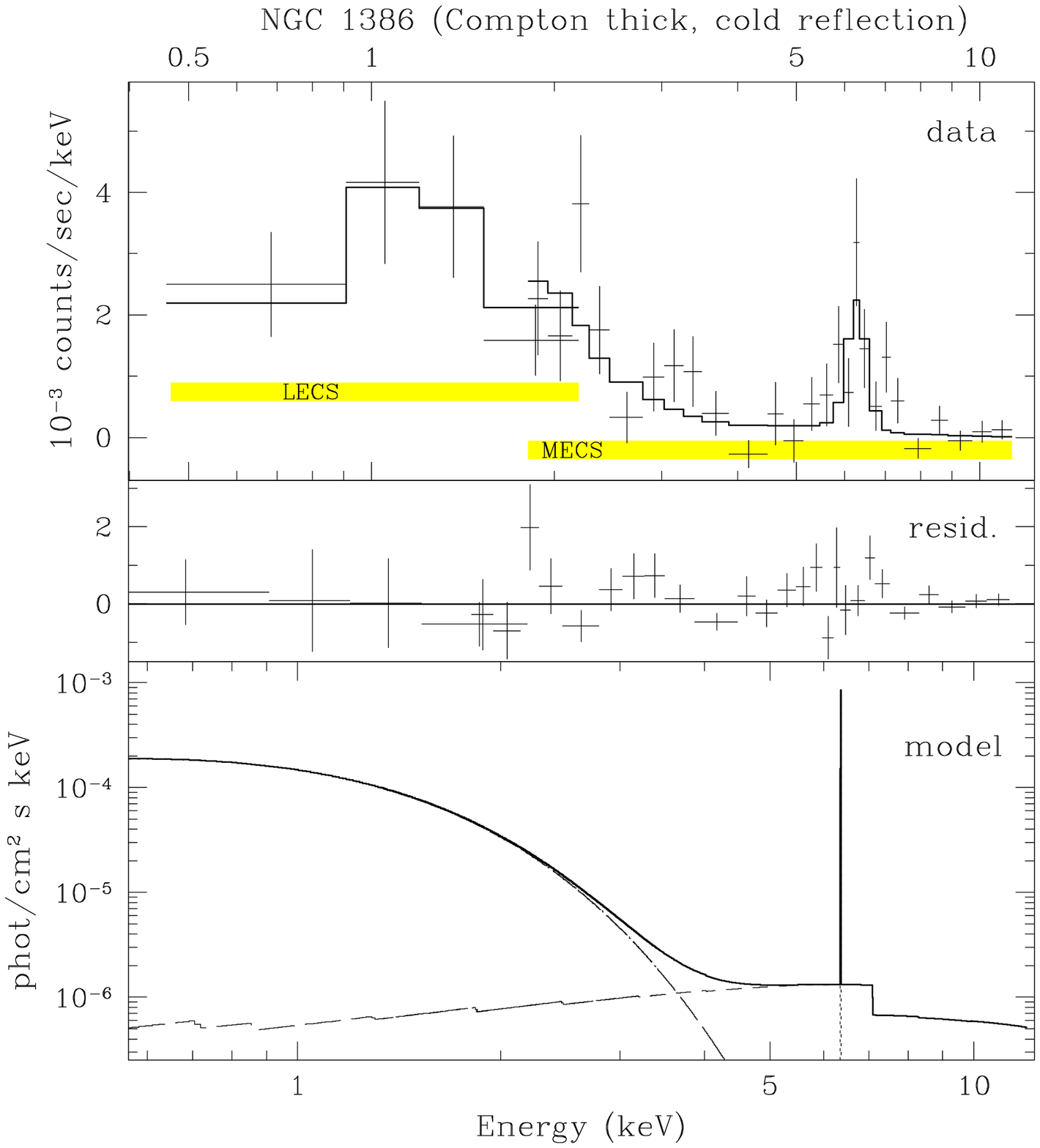} \includegraphics{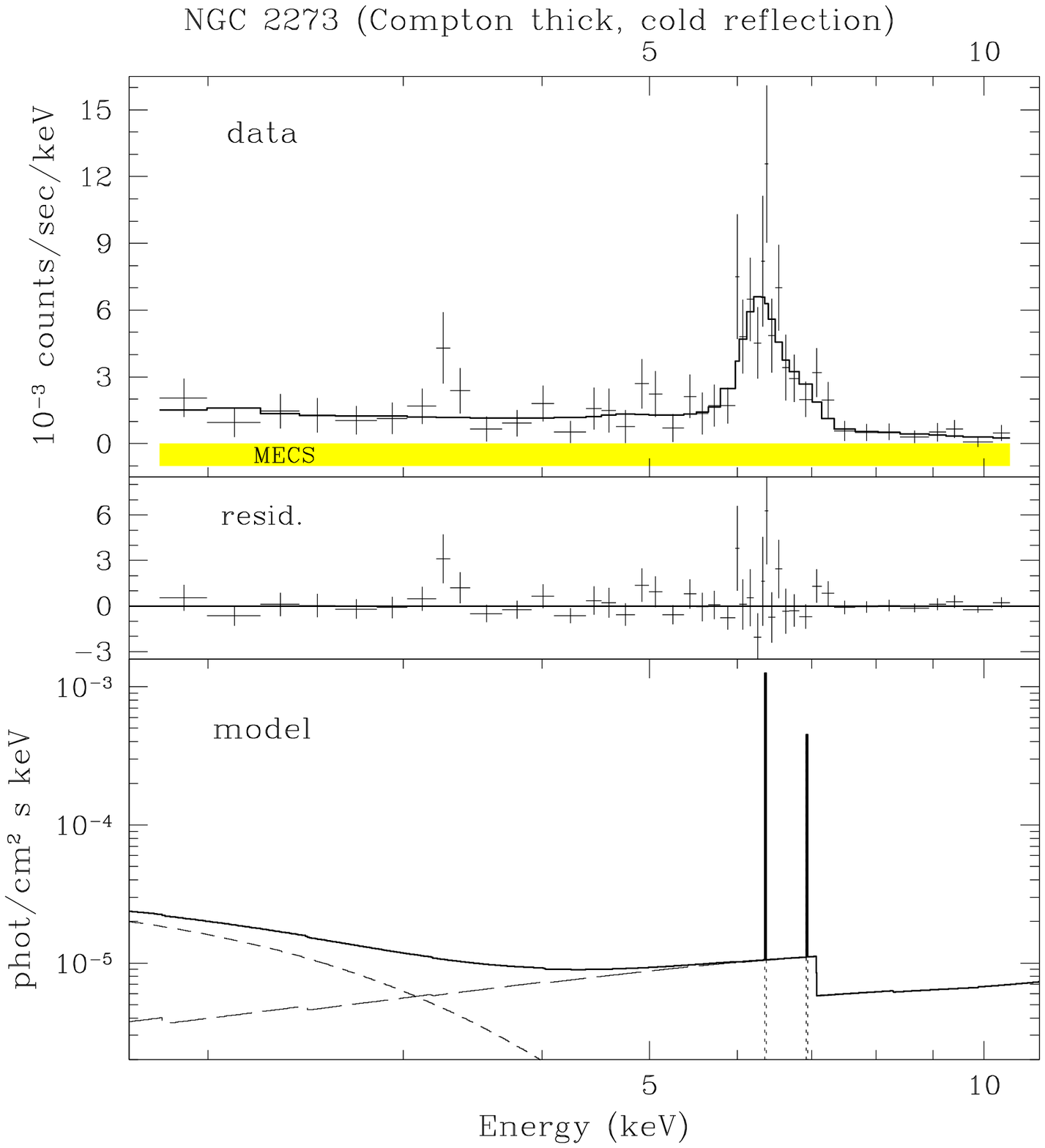}}
\resizebox{\hsize}{!}{\includegraphics{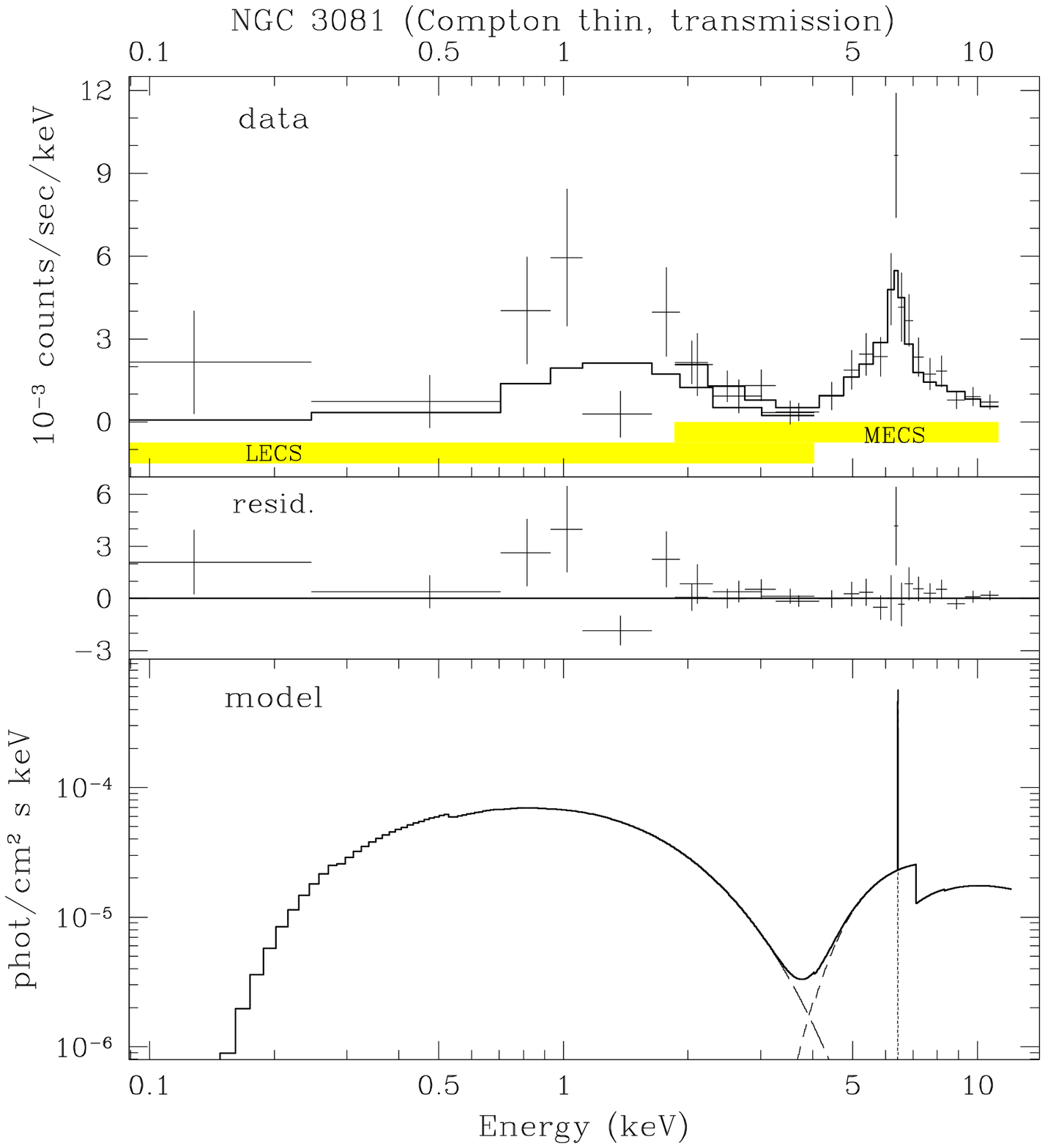} \includegraphics{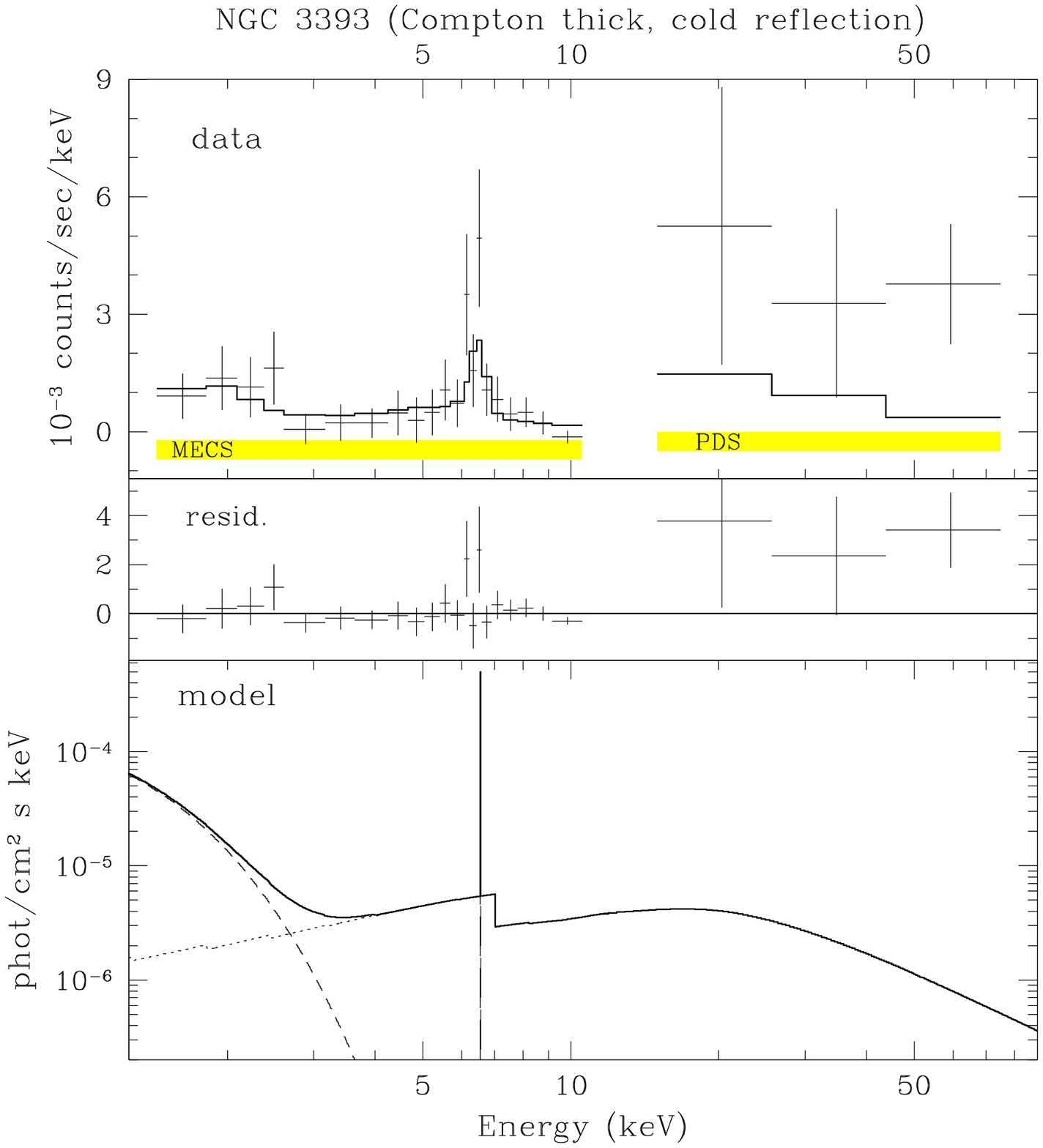}}
\parbox[b]{\hsize}{
 \caption{Each box shows the data and the folded best fit
model (top), the residuals
(middle) and the unfolded model (bottom)
 for four of the objects in our sample. The shaded regions in the top
panels indicate the energy bands selected for each instrument.}
\label{fig_spectra_a}}
\end{figure*}
\begin{figure*}[!]
\resizebox{\hsize}{!}{\includegraphics{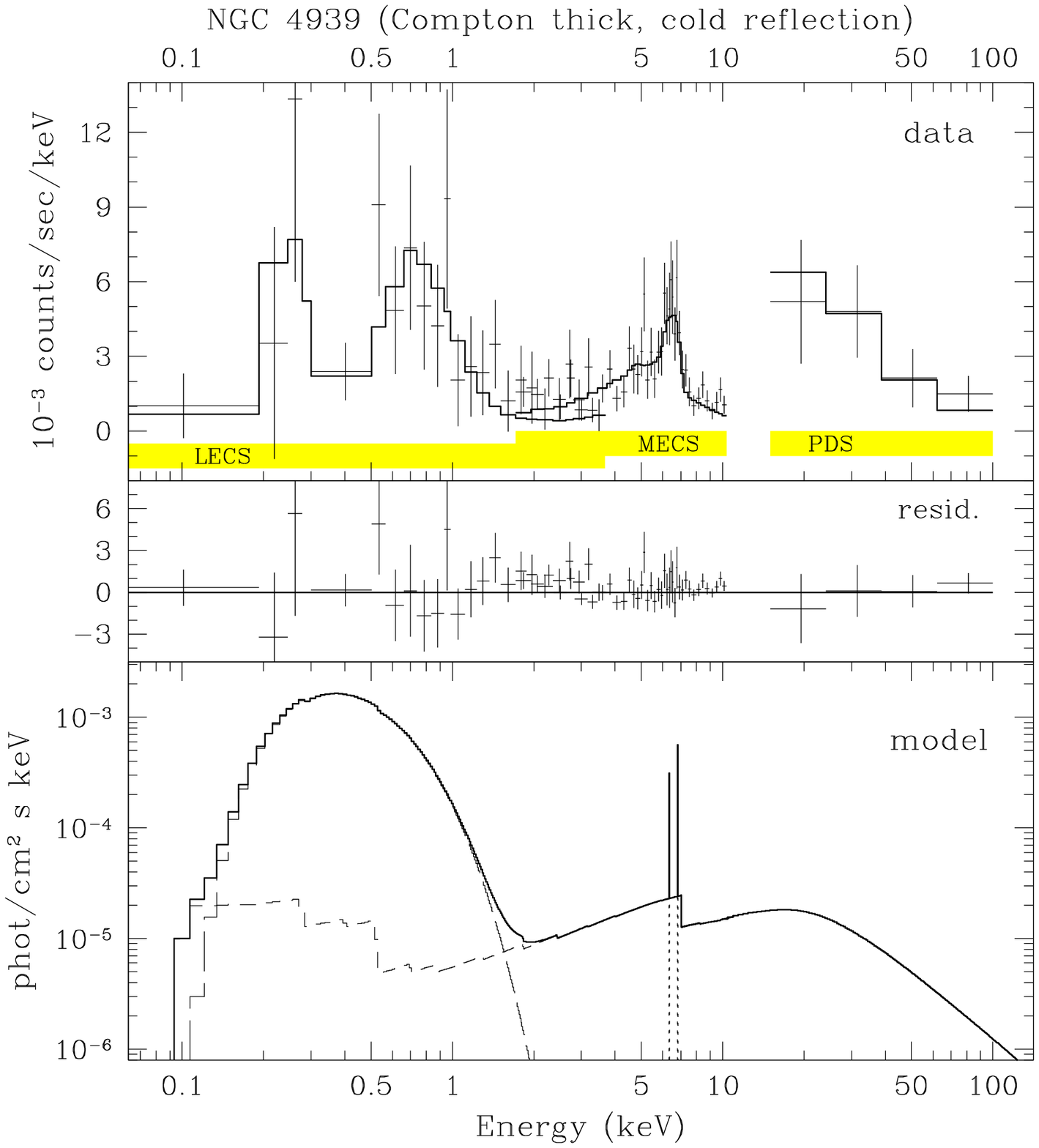} \includegraphics{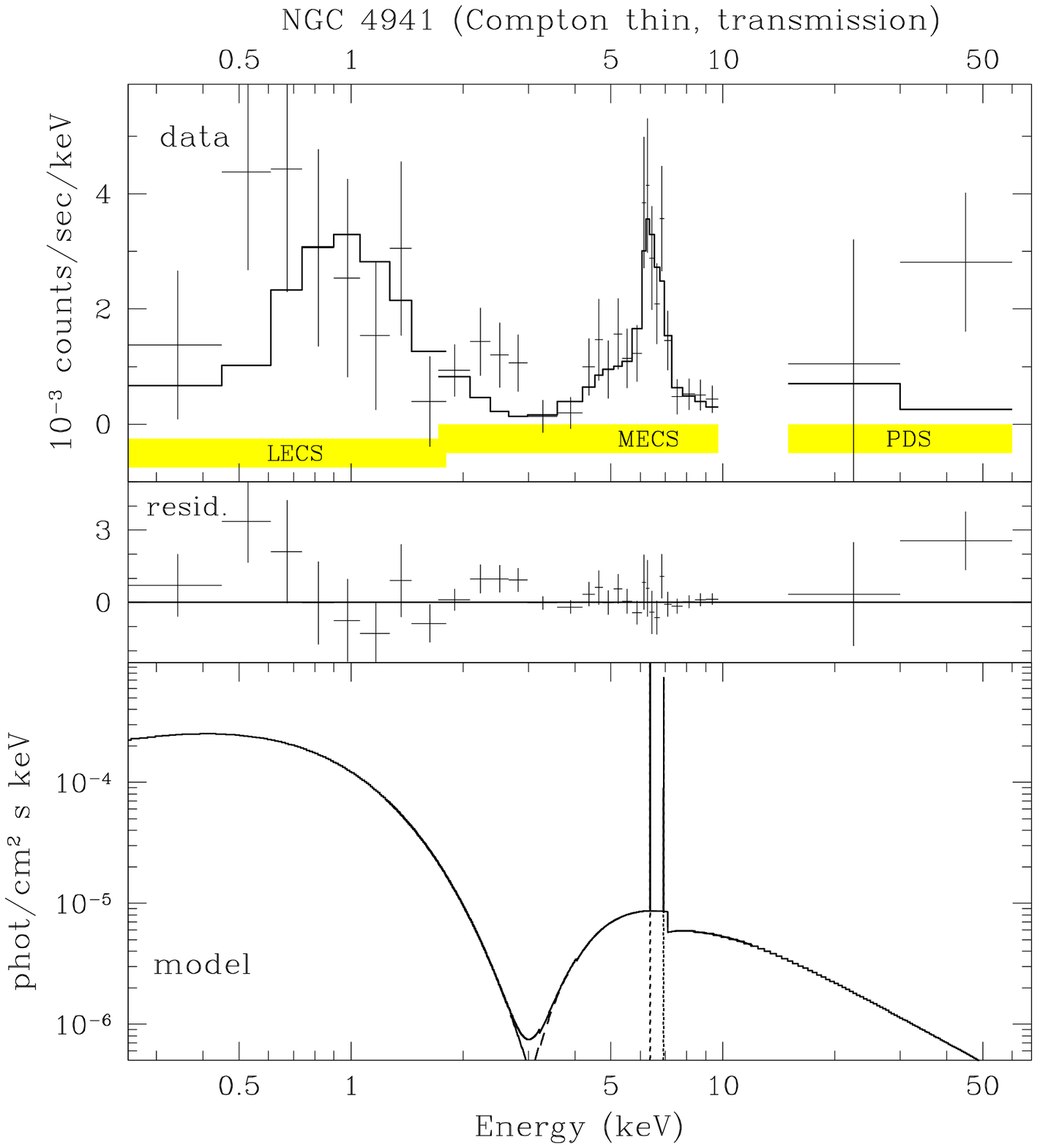}}
\resizebox{\hsize}{!}{\includegraphics{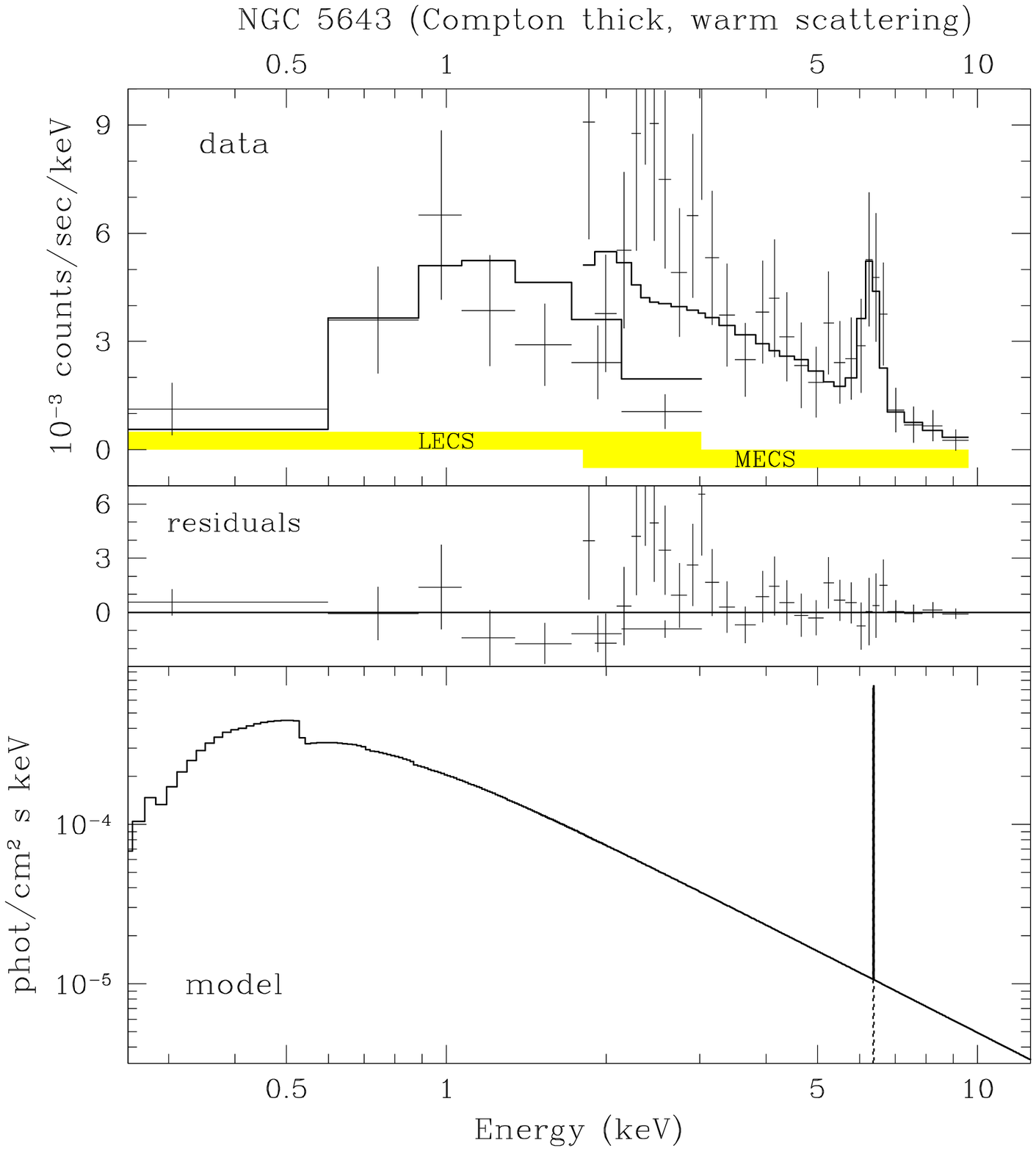} \includegraphics{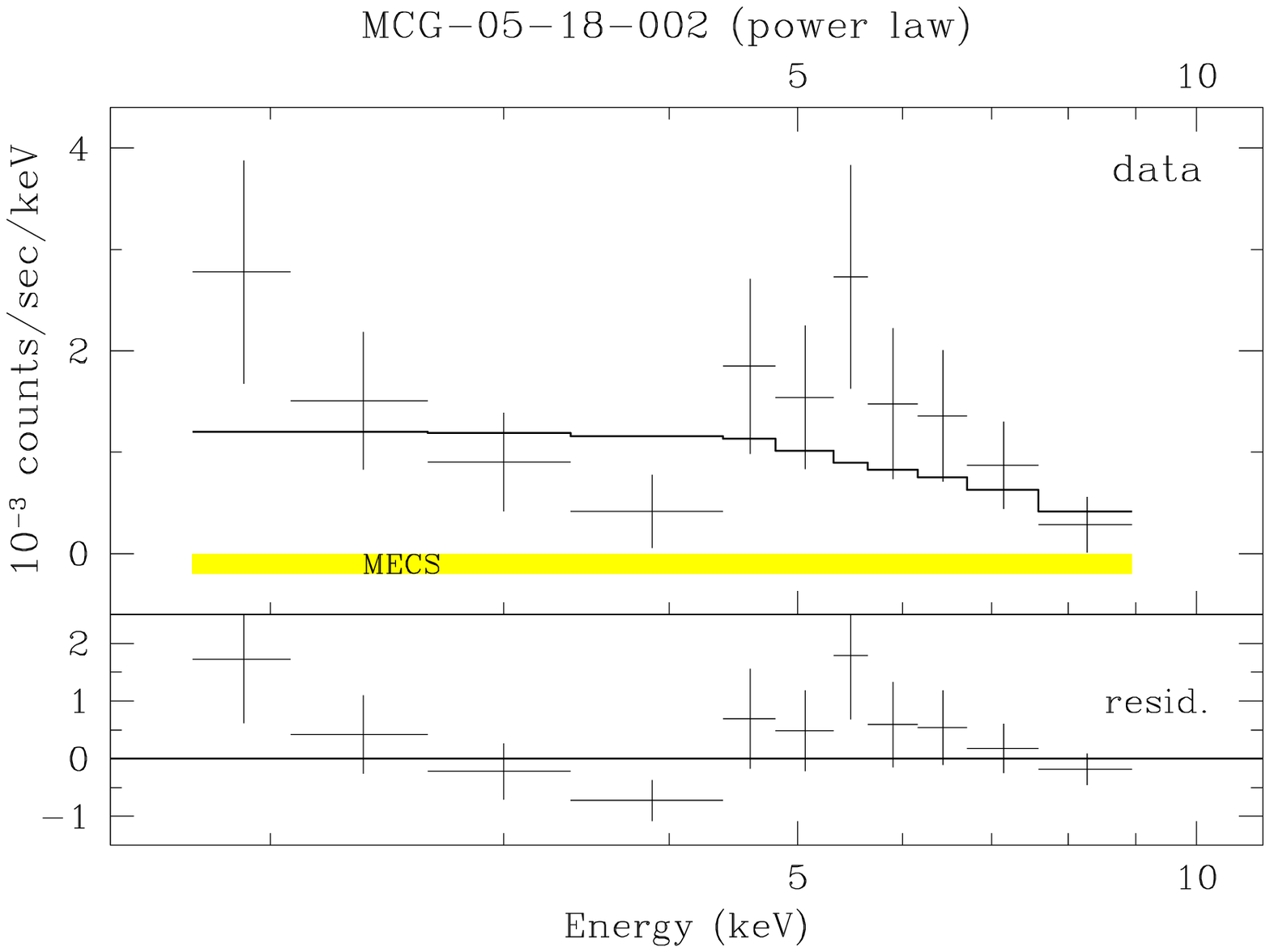}}
\parbox[b]{\hsize}{
 \caption{Each box shows the data and the folded best fit
model (top), the residuals (middle) and the unfolded model (bottom)
for four of the objects in our sample. The shaded regions in the top
panels indicate the energy bands selected for each instrument.}
\label{fig_spectra_b}}
\end{figure*}

\begin{table*}
\rotcaption[]{Spectral fits} \label{tab_results}
\begin{sideways}
\vbox{\vspace{-0.5cm}
\hbox{ \hspace{1.8truecm}
\vspace{-0.8truecm}
\begin{tabular}{lcccccccc}
\vspace{.01truecm}\\
\multicolumn{9}{c}{Cold reflection dominated models (Compton thick,
  N$_H >$10$^{25}$cm$^{-2}$)$^a$} \\
\hline
Source & kT &  \multicolumn{3}{c}{Fe Line} &
F$_{0.5-2keV}$ &
F$_{2-10keV}$ & F$_{20-100keV}$ & $\chi ^2$/d.o.f.\\
       &  &  Energy$^b$ & Normaliz. & EW & 
 & & & \\
       & (keV) &  (keV) & (10$^{-5}$ph/cm$^2$/s) & (keV) &
    \multicolumn{3}{c}{(10$^{-13}$erg/cm$^2$/s)} & \\
\hline
\bf *NGC 1386 & \bf 0.36 &  $\bf 6.38^{+0.14}_{-0.13}$ &
  $\bf 1.00^{+0.43}_{-0.47}$ & $\bf 7.6^{+8.9}_{-5.0}$ & 
 \bf 2.87 & \bf 2.40 & -- & \bf 33.2/38 \\
\bf *NGC 2273 & \bf 0.58 &  $\bf 6.4^c$ & $\bf 2.36^{+1.35}_{-0.70}$ &
  $\bf 2.49^{+0.80}_{-0.68}$ & -- & \bf 9.93 & -- & \bf 30.7/39 \\
 &  &  $\bf 6.96^a$ & $\bf 0.88^{+0.59}_{-0.57}$ & $\bf 0.94^{+0.59}_{-0.57}$
  &  & & & \\
\bf *NGC 3393 & \bf 0.30 &  $\bf 6.64^{+0.20}_{-0.32}$ &
  $\bf 0.96^{+0.74}_{-0.60}$ & $\bf 1.89^{+3.8}_{-1.2}$ & -- & \bf 3.9 & \bf 54
  & \bf 20.1/19 \\
\bf *NGC 4939 & \bf 0.14 & $\bf 6.4^c$ & $\bf 0.97^{+1.0}_{-0.38}$ &
  $\bf 0.48^{+0.42}_{-0.21}$  & \bf 3.92 & \bf 14.2 & \bf 234 & \bf 83.0/94 \\
 &  &  $\bf 6.96^a$ & $\bf 0.69^{+0.88}_{-0.33}$ & $\bf 0.30^{+0.43}_{-0.14}$
  &  & & & \\
NGC 4941 & 0.19 & 6.4$^c$ & 1.20$^{+0.53}_{-0.59}$ & 1.35$^{+0.78}_{-0.89}$  &
  2.3 & 6.23 & -- & 34.3/47 \\
 &  &  6.96$^a$ & 0.89$^{+0.49}_{-0.54}$ & 1.35$^{+0.78}_{-0.89}$ &  & & & \\
NGC 5643 & 0.73 &  6.46$^{+0.20}_{-0.14}$ & 1.75$^{+1.07}_{-0.90}$ &
   1.41$^{+0.93}_{-0.73}$ & 2.32 & 13.5 & -- & 35.6/31 \\
\hline
\hline
\end{tabular}
}}
\end{sideways}
\begin{sideways}
\vspace{.5truecm}\\
\hbox{ \vspace{0.2truecm} \hspace{1.8truecm}
\begin{tabular}{lccccccccc}
\vspace{.5truecm}\\
\multicolumn{9}{c}{Warm scattering dominated models (Compton thick,
  N$_H >$10$^{25}$cm$^{-2}$)$^a$} \\
\hline
Source & kT &  \multicolumn{3}{c}{Fe Line} & F$_{0.5-2keV}$ &
F$_{2-10keV}$ & F$_{20-100keV}$ &  $\chi ^2$/d.o.f.\\
       &  &  Energy$^b$ & Normaliz. & EW & 
 & & & \\
       & (keV) & 
         (keV) & (10$^{-5}$ph/cm$^2$/s) & (keV) &
    \multicolumn{3}{c}{(10$^{-13}$erg/cm$^2$/s)} & \\
\hline
NGC 1386 & 0.35 & $6.38^{+0.18}_{-0.17}$ & $1.09^{+0.50}_{-0.57}$ &
 $>3.3$ & 3.15 & 2.18 & -- &35.3/38 \\
NGC 2273 & -- & $6.4^c$ & $3.12^{+0.58}_{-0.80}$ & $8.3^{+1.4}_{-2.3}$ &
 -- & 9.54 & -- & 36.2/41 \\
  & & $6.96^a$ & $1.26^{+0.52}_{-0.55}$ & $3.9^{+1.3}_{-1.8}$ & & & & \\
NGC 3393 & 0.22 & $6.65^{+0.15}_{-0.30}$ & $1.37^{+0.72}_{-0.70}$ &
 $8.3^{+12}_{-4}$ & -- & 3.12 & 2.6 & 24.7/19 \\
NGC 4941 & -- & $6.4^c$ & $1.63^{+0.54}_{-0.54}$ & $6.2^{+3.6}_{-1.8}$ &
 1.75 & 5.42 & -- & 39.6/47 \\
  & & $6.96^a$ & $1.12^{+0.64}_{-0.41}$ & $4.9^{+4.3}_{-1.6}$ & & & & \\
\bf *NGC 5643 & -- &  $\bf 6.46^{+0.20}_{-0.12}$ & $\bf 2.0^{+1.2}_{-0.8}$ & 
 $\bf 1.9^{+1.4}_{-0.7}$ & 
 \bf 4.48 & \bf 11.2 & -- & \bf 32.4/33\\
\hline
\hline
\end{tabular}
}
\end{sideways}
\begin{sideways}
\begin{tabular}{lcccccccccc}
\vspace{-0.3truecm}\\
\multicolumn{11}{c}{Transmission models (Compton thin)} \\
\hline
Source & kT & N$_H$ & \multicolumn{4}{c}{Fe Line} &
F$_{0.5-2keV}$ & F$_{2-10keV}$ & F$_{20-100keV}$ & $\chi ^2$/d.o.f.\\
   &   & & Energy$^b$ & Normaliz. & EW & (EW$_{corr}$) & 
 & & \\
       & (keV) & (10$^{23}$cm$^{-2}$) &
         (keV) & (10$^{-5}$ph/cm$^2$/s) & \multicolumn{2}{c}{(keV)} &
    \multicolumn{3}{c}{(10$^{-13}$erg/cm$^2$/s)} & \\
\hline
NGC 1386 & 0.37 & $8^{+\infty}_{-8.}$ & $6.38^{+0.18}_{-0.17}$ & 
 $ 0.95^{+0.45}_{-0.46}$ & $4.7^{+9.2}_{-2.3}$ &
 $ 1.16^{+0.56}_{-0.56}$ &  2.60 & 2.60 & -- & 33.12/37 \\
\bf *NGC 3081 & \bf 0.46 & $\bf 6.4^{+2.0}_{-1.2}$ & $6.48^{+0.67}_{-0.16}$ & 
 $\bf 1.33^{+1.02}_{-0.80}$ & $\bf 0.57^{+0.44}_{-0.34}$ &
 $\bf 0.20^{+0.20}_{-0.12}$ & \bf 1.51 & \bf 13.3 & -- & \bf 40.8/37\\
NGC 3393 & 0.36 & $3.2^{+10.}_{-2.5}$ & $6.70^{+0.19}_{-0.41}$ & 
 $ 0.96^{+0.67}_{-0.74}$ & $2.0^{+1.4}_{-1.5}$ &
 $ 1.4^{+0.8}_{-1.0}$ &  -- & 3.85 & 12.5 & 21.2/18 \\
NGC 4939 & 0.40 & $3.0^{+2.9}_{-1.8}$ & $6.4^c$ &
 $ 1.0^{+0.8}_{-0.6}$ & $0.49^{+0.41}_{-0.29}$ &
 $ 0.26^{+0.22}_{-0.15}$ &  2.05 & 15.0 & 61.5 & 123/93 \\
  &  &  & $6.96^c$ & $ 1.1^{+0.6}_{-0.6}$ & $0.53^{+0.30}_{-0.29}$ &
 $ 0.28^{+0.16}_{-0.15}$ &  &  &  &  \\
\bf *NGC 4941 & \bf 0.25 & $\bf 4.5^{+2.5}_{-1.4}$ & $\bf 6.4^a$ &
  $\bf 1.2^{+0.5}_{-0.6}$ & $\bf 1.6^{+0.7}_{-0.9}$ & 
  $\bf 0.72^{0.32}_{-0.40}$ & \bf 2.08 & \bf 6.63 & -- & \bf 34.9/46 \\
 & & & $\bf 6.96^a$ & $\bf 0.81^{+0.53}_{-0.54}$ & $\bf 0.9^{+0.6}_{-0.6}$ & 
  $\bf 0.56^{0.37}_{-0.37}$ &  &  & &  \\
\hline
\hline
\end{tabular}
\end{sideways}
\begin{sideways}
\vbox{\vspace{0.2cm}
\begin{minipage}{20cm}
In these models the photon index of the primary radiation is frozen to
1.7. Bold face entries (also marked with an asterix)
indicate sources for which the model provides
the best fit, both in terms of $\chi ^2$ and in terms of EW(Fe) and
L$_X$/L$_{[OIII]}$ properties (see text). Spectral models that are 
inconsistent with the data at a high significance level are not reported.
MCG-05-18-002 was not fitted with any of these models because of the low
signal--to--noise; for this source we only report a simple power law fit
in Table~\ref{tab_apx}. The cold reflection spectrum was modeled with
the XSPEC routine PEXRAV.
\\ Notes:
$^a$ In the case of NGC1386 the Compton thick model constrains only N$_H >$
  10$^{24}$cm$^{-2}$ (see text);
$^b$ rest frame;
$^c$ frozen parameter.
\end{minipage}}
\end{sideways}
\end{table*}

\subsection{Results on single objects} \label{results}

In this section we describe the results of spectral fits for the objects in
our sample. Also,
we apply the considerations discussed in the former sections
to each single object.

Table~\ref{tab_results} shows the parameters of the three fitting models
for each of the sources in our sample. Bold face entries (also marked with
an asterix) indicate sources
for which the corresponding model provides the best fit, according to
the considerations discussed below.
Instead, spectral fits that are inconsistent with the data
at a high significance level are not reported.
The second column (kT) indicates the temperature of the black body
used to fit the soft excess. As for the Fe line we report line energy,
normalization and equivalent width. In the transmission models we also
report the EW(Fe) once the underlying continuum is corrected for the
absorbing column density N$_H\ $, listed in column 3 (EW$_{corr}$):
this value provides a
lower limit to the absorption corrected EW(Fe), since it does not take
into account the absorption affecting the Fe line itself;
if this lower limit turns out to be higher than about 500 eV (i.e.
significantly higher than the $\sim 200$ eV observed in Sy1s) this would
indicate that the Compton thin model is little plausible on physical grounds.
All errors are at the 90\% confidence level for one interesting parameter.
The last column indicates the $\chi ^2$/{\it degrees of freedom}
 for the whole fit,
i.e. including the low energy data and the black body component. However,
when discussing various spectral models in the following sections, we will often
refer to  $\chi ^2$ differences relative to the high energy ($>$3keV) data
alone.

Figures \ref{fig_spectra_a} and \ref{fig_spectra_b}
show the data along with the folded best fit models,
the residuals from the model and the
unfolded models. Hereafter we discuss each object individually.

\subsubsection{NGC 1386}

This source was detected also in the 20--100 keV spectral region by the PDS.
However, further analysis indicated that most of the flux is due to a
nearby Seyfert galaxy (NGC1365) that happens to be located at the edge
of the PDS beam. As a consequence, we could not use the PDS data to constrain
the X--ray properties of the source.

The MECS spectrum is characterized by a prominent iron line
and a very low continuum level. The low continuum makes estimates of the
EW(Fe K$\alpha$) very uncertain, since the latter becomes very sensitive to the
continuum fit. However, we could set a 90\% confidence lower limit of 
$\sim$ 2 keV to the EW(Fe K$\alpha$),
that strongly supports the Compton thick nature of this source.

The Compton, cold reflection
model provides the best fit to the high energy ($>$~3~keV) data. However, from 
a statistical point of view this model is only marginally better than the warm
scattering model: $\Delta \chi ^2 = 1$ with the same number of degrees of 
freedom. The main problem is that the low continuum level makes difficult
discriminating different continuum shapes.

On the other hand, the Fe K$\alpha$
line center is consistent with 6.4 keV, and it is not consistent with
6.7 or 6.96 keV at a high confidence level.
There is no evidence for additional components at 6.7 or 6.96
keV. This result provide further support to the cold reflection model
with respect to the warm scattering model.

As discussed above, the PDS data cannot be used to constrain the
emission in the 20--100 keV range of NGC 1386.
As a consequence, in Compton thick models
we cannot rule out an absorbing colummn density in the range
$10^{24}-10^{25}$cm$^{-2}$ that would provide an excess of emission in the
PDS band with respect to the MECS flux. Therefore,
in Compton thick models we can only determine a lower limit of
$10^{24}$cm$^{-2}$ for the absorbing column density.

In the Compton thin, transmission model the absorbing column 
density N$_H$ is indetermined. A value of
N$_H \approx 8\times 10^{23}$ cm$^{-2}$ provides a fit that
 statistically is not significantly worse than the
cold reflection model. The Compton thin model is ruled out
on physical grounds because of the large EW(Fe K$\alpha$). Even the EW computed
by correcting the continuum for absorption (EW$_{corr}$) has still a 90\%
lower limit of 600 eV.

\begin{figure}[!]
\resizebox{\hsize}{!}{\includegraphics{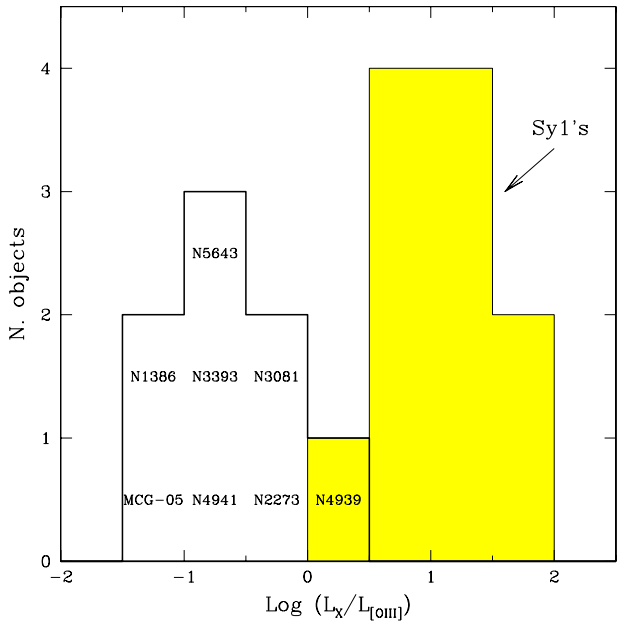}} 
 \caption{Distribution of the ratio between the observed 2--10 keV luminosity
and the (reddening corrected) [OIII] line luminosity
 for the sources in our sample,
compared to the same distribution for the Sy1s in the sample of Mulchaey et
al. (\cite{mulchaey}) (we excluded the Sy1s for which the
narrow-lines Balmer decrement was not available to correct the [OIII] flux).}
\label{fig_xoiii}
\end{figure}

The Compton thick model 
is also supported by the L$_X$/L$_{[OIII]}$ ratio
which, as shown in Fig.~\ref{fig_xoiii}, is about two orders of magnitude
lower than in Sy1s.

NGC 1386 has been observed also by ASCA on 1995 January 26, with a similar
integration time as our SAX observation. We reduced and analyzed the ASCA
data and found that both flux and spectral shape
of the ASCA spectrum are consistent with our data within the uncertainties.
This indicates that 
the source does not show evidence for long term variability in excess of
about 25\% , that is our uncertainty on the flux. This is consistent
with the idea that the observed flux is not seen directly, but is reprocessed
by a large scale ($\ge$ 1 pc) medium.

The ASCA data of NGC 1386 were also analyzed by Iyomoto et al. (\cite{iyomoto})
 who interpret the observed spectrum with a Compton thin transmission model
(N$_H\approx 2.8-5.4\times 10^{23}$cm$^{-2}$). As discussed above,
statistically the transmission model would fit also our data, 
but is inconsistent with the large EW(Fe K$\alpha$).

Finally, we estimated the contribution
to the soft X~rays from the Fornax cluster thermal emission by extracting
the spectrum in two regions of the sky located at the same distance from
the cluster center, and by using the same aperture size. The contribution
from the cluster to the observed soft X--ray flux turns out to be about one
third of the total observed in NGC 1386.

\subsubsection{NGC 2273}

The Compton thick, cold reflection model fits the high energy data better than
the warm scattering model at a high confidence level: above 3 keV,
$\Delta \chi ^2 = 6$ with the same number of degrees of freedom.
The Compton thin, transmission model indicates N$_H = 2.6\times 10^{23}$
cm$^{-2}$ and is worse than the cold reflection model only at the
70\% confidence level (same $\chi ^2$ with only one degree of freedom less).
However, the Compton thick nature of this spectrum
is supported by the large EW of the iron line(s). Also, if 
in the transmission model the photon index is thawed (see Appendix) the best
fitting value is $\Gamma = 0.7$,
much flatter than expected for the intrinsic
spectrum, thus further supporting the Compton thick, cold reflection model.

In Compton thick models the upper limit provided by the PDS in the
20--100 keV range rules out an absorbing column density in the range
$10^{24}-10^{25}$cm$^{-2}$, otherwise the observed 20-100 keV
flux would be significantly higher than our upper limits. Therefore,
if the Compton thick model is valid the column density along our
line of sight must be larger than $10^{25}$cm$^{-2}$.

The fit of the iron line improves significantly ($>$99\% confidence
level) by splitting the line in two components at 6.4 and 6.96 keV.
The bump at 3.1 keV observed in the residuals could be interpreted as
a line of ArXVII, but by adding a gaussian at this location the fit
does not improve significantly.

\subsubsection{NGC 3081}

The Compton thin, transmission model (N$_H = 6.4 \times 10^{23}$ cm$^{-2}$)
fits the high energy data better than the Compton thick models at a high 
confidence level (the latter ones have
reduced $\chi _r^2 > 2$). The absorbing column density is very large, and 
is responsible for making the observed EW(Fe K$\alpha$) intermediate 
between typical Compton thin sour\-ces and Compton thick ones.

\subsubsection{NGC 3393}

The MECS data of this source were already presented in Salvati et al.
(\cite{salvati}),
in this paper we include the PDS detection in the 20--100 keV range.

As discussed in Salvati et al. (\cite{salvati}), statistically
the Compton thick, cold reflection model is only marginally better than the
transmission model: the latter (N$_H = 3.7\times 10^{23}$cm$^{-2}$)
fits the data with a $\Delta \chi ^2 =1$ with respect to the former,
at the expense of only one extra degree 
of freedom. However, the data above 10 keV are reasonably consistent
with cold reflection, while the transmission
model falls short of accounting for the high energy data.
If in the transmission model the photon index is thawed (see Appendix) the best
fitting value of $\Gamma$
is negative, that is certainly not acceptable for the intrinsic
spectrum.
The EW(Fe K$\alpha$) is large even when the continuum is corrected for the
absorbing column deduced by the transmission model. Both findings are
in favor of the Compton thick interpretation (with N$_H > 10^{25}$ cm$^{-2}$).
The warm scattering model is ruled out at a confidence level larger than
90\% and, in particular, falls short of accouting for the PDS flux.

\subsubsection{NGC 4939}

The Compton thick, cold reflection model (with N$_H>10^{25}$cm$^{-2}$,
i.e. reflection throughout the PDS range) provides the best fit to the data
in the 3--100 keV spectral range. The warm scattering model is ruled
out at a high significance level (reduced $\chi _r^2 > 2$); 
the Compton thin, transmission model (N$_H=3.0\times 10^{23}$
cm$^{-2}$)
is worse than the cold reflection model at a confidence level larger than 90\%
(above 3 keV, $\Delta \chi ^2 = 6$ with only one degree of freedom less); 
both alternatives are especially bad in the 10--100 keV band, where they 
fall short of accounting for the observed flux. As in the case of NGC2273,
if in the transmission model the photon index is thawed (see Appendix) the best
fitting value is 0.7, adding
further support to the Compton thick, cold reflection interpretation.

The iron line appears to contain a H--like component at 6.96 keV
that is as strong as the neutral-fluorescence line at 6.4 keV. Perhaps a 
warm scattered component is present at lower energies (i.e. NGC 4939
might be similar to NGC 1068, Matt et al. \cite{matt_d}),
but the signal--to--noise
in our data is not high enough to detect any additional component of
the continuum.

We should note that the EW of the 6.4 keV line is only 400 eV
and that the L$_X$/L$_{[OIII]}$ only about one order of magnitude
lower than the average in Sy1s (Fig.~\ref{fig_xoiii}).
As a consequence, the preference for the cold reflection, Compton thick
model over a transmission model with N$_H=3\times10^{23}$cm$^{-2}$ 
is still questionable.

\subsubsection{NGC 4941}

The MECS data of NGC 4941 were also
presented in Salvati et al. (\cite{salvati}). In this
paper we include the LECS data and a marginal PDS detection in the 30--60 keV
range. However, the latter datum was not used during the fitting
procedure because of its low statistical significance (though it is plotted
in Fig.~\ref{fig_spectra_b}).
The transmission model (N$_H =4.5\times 10^{23}$cm$^{-2}$)
provides the best fit to the data above 3 keV.
The Compton thick warm scattering model is ruled out at a high confidence level,
and the Compton thick, cold
reflection model is also worse than the transmission model
(though at a lower confidence level: above 3 keV $\Delta \chi ^2 =4$
with only one
degree of freedom less). The EW(Fe$_{6.4keV}$) is high, but 
consistent with 300 eV at the 90\% level once the effects of absorption 
are taken into account. Summarizing, although the transmission
scenario is preferred, we consider the nature of this spectrum still
uncertain.

\subsubsection{NGC 5643}

This source is close to a cluster of galaxies. To avoid contamination 
we reduced the extraction radius to 1$'$.
We also checked the level of contamination by extracting the spectrum of
two regions located at the same distance from the cluster as NGC5643
and found that the flux from the cluster
contributes a negligible fraction of the total flux observed in NGC5643.

A simple power law with $\Gamma = 1.7$, and no excess absorption
with respect to the Galactic value (N$_H=8\times 10^{20}$cm$^{-2}$)
fits the continuum significantly better than the cold reflection model.
Both the
absence of an absorption cutoff, in addition to the Galactic one, and
the large EW of the iron line strongly support the idea
that the source is Compton thick, and warm scattering dominated.

If this were the case, ionized iron lines would be expected.
The best fit energy of the gaussian in the rest frame of the
galaxy is 6.46$^{+0.20}_{-0.14}$, i.e. the iron is less
ionized than Fe {\sc xxii}.
These ions can still emit a significant line by fluorescence and
resonant scattering; lighter elements are highly ionized so that 
the scattered spectrum is not dramatically distorted in shape with
respect to the incident one.

NGC 5643 has been observed also by ASCA for $\sim 40$ ksec on 1996
February 21. The 2--10 keV flux observed by ASCA is about 30\% lower 
than that derived from our data, while there is agreement on the EW 
and the energy of the Fe line.
Above 3 keV the slope of the power law in the ASCA spectrum
is flatter than that observed in the BeppoSAX spectrum. The difference in
flux is puzzling, calibration problems might have occurred.
The difference in slope could be ascribed to the low statistics in both
spectra (actually an intermediate model could fit them both with
a reduced $\chi_r^2 \sim 1$.), but it calls into question our interpretation
of the spectrum as warm scattering dominated. However, be it warm scattering
or cold reflection dominated, the Compton thick nature of the source is 
supported by both ASCA and BeppoSAX data.

As discussed previously, the upper limit provided by the PDS 
requires an absorbing column density higher than $10^{25}$cm$^{-2}$.

\subsubsection{MGC-05-18-002}

This source was observed after the failure of one of the three MECS units.
Both the reduced effective area and the short integration time are
responsible for the low signal--to--noise ratio of its spectrum.

In this case we only used a simple power law fit (Table~\ref{tab_apx}).
Although a
detailed analysis cannot be performed, both the very flat spectrum
(Tab.~\ref{tab_apx}) and the very low L$_X$/L$_{[OIII]}$ ratio (two orders
of magnitude lower than the average Sy1, as shown in Fig.~\ref{fig_xoiii})
strongly
suggest that this is another Compton thick source, probably cold reflection
dominated, and with N$_H>10^{25}$cm$^{-2}$ because of the
lack of detection in the PDS range.
Higher signal--to--noise data are required to confirm this
interpretation.

\section{Discussion} \label{discussion}

\subsection{Probing X--ray weak AGNs} \label{disc_lla}

\begin{figure}[!]
\resizebox{\hsize}{!}{\includegraphics{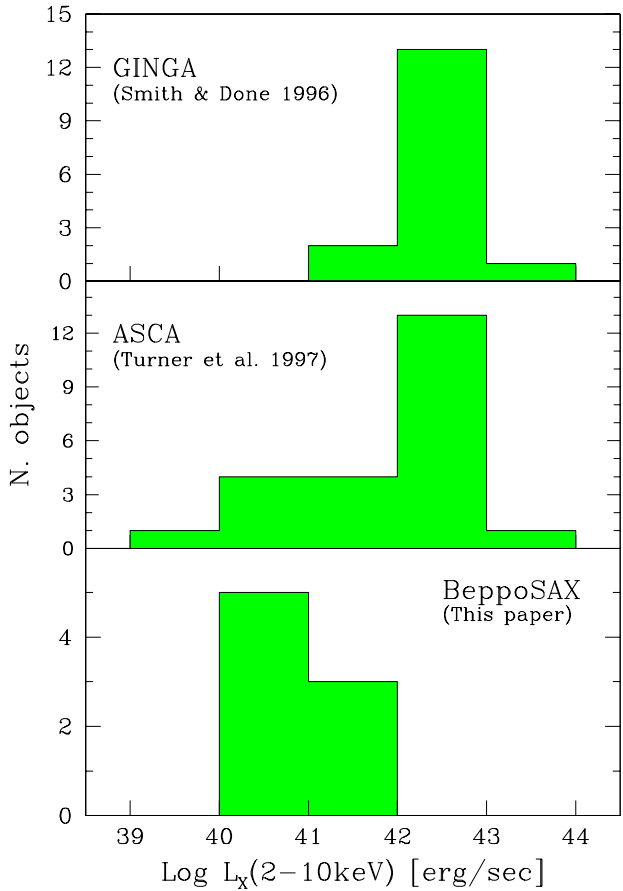}} 
 \caption{Distribution of the observed luminosities in the 2--10 keV band 
of our sample (bottom) compared to the same distribution in other
hard X--ray spectroscopic surveys of Sy2s made with GINGA (top)
and ASCA (middle).}
\label{fig_lum}
\end{figure}

As discussed in Sect.~\ref{intro} most of the former hard X--ray spectroscopic
surveys of Sy2s were seriously biased for X--ray bright sources.
This is illustrated in Fig.~\ref{fig_lum}.
The observed 2--10 keV luminosity in the Sy2 (and Sy1.9) galaxies
surveyed by GINGA, most of which are reported in Smith \& Done (\cite{smith}),
has an average $\langle log~L_{2-10keV}(erg/s)\rangle =42.5$.
Turner et al. (\cite{turner_a},\cite{turner_b})
studied lower luminosity Sy2s by using ASCA 
data: their sample has a mean $\langle log~L_{2-10keV}(erg/s)\rangle =41.8$,
but it includes some objects with luminosity
lower than $10^{41}$ erg/sec.

Our sample was not selected according to the X--ray flux. Also, we avoided
objects for which hard X--ray spectroscopic data were already available. As a
consequence, our survey samples Sy2s with X--ray fluxes
significantly lower than in former studies, although at similar distances. 
Indeed, in our sample  $\langle log~L_{2-10keV}(erg/s)\rangle =40.8$.

\begin{figure}[!]
\resizebox{\hsize}{!}{\includegraphics{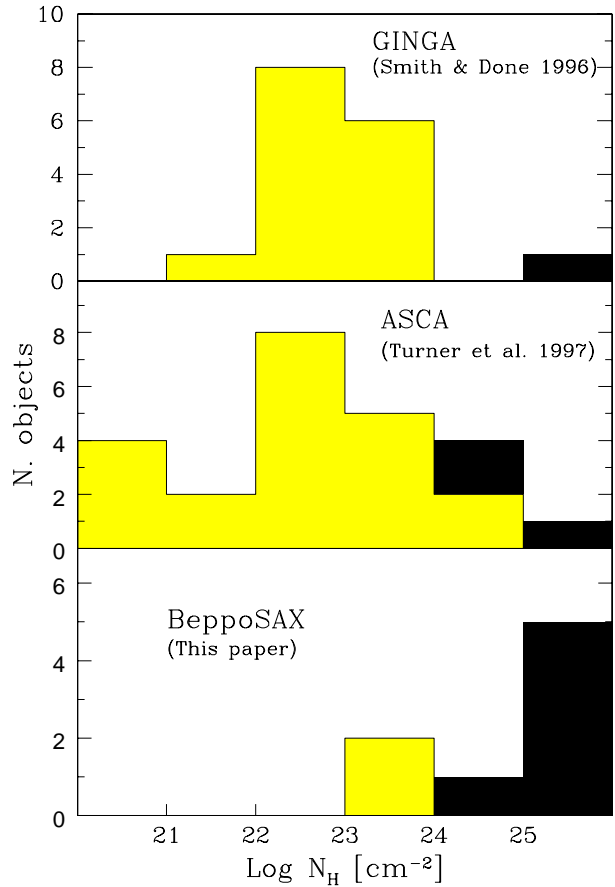}} 
 \caption{Distribution of the absorbing column density N$_H$
derived for the Sy2s in our
sample (bottom) compared to the same distribution in other
hard X--ray spectroscopic surveys of Sy2s made with GINGA (top)
and ASCA (middle).
The black histograms indicate Compton thick
objects for which only lower limits on N$_H$ could be determined.}
\label{fig_nh}
\end{figure}

These X--ray weak AGNs have spectral properties similar to the
bright ones: a
relatively flat continuum and the presence of the Fe K$\alpha$ emission
line blend, generally dominated by the neutral component at 6.4 keV. 
The main difference between the X--ray properties of the
sources in our sample and bright AGNs is that the former
have {\it observed} spectra that are generally flatter (see Appendix)
and are characterized by Fe lines with larger EW;
as discussed in Sect.~\ref{results},
these differences are very likely due to heavy obscuration
along our line of sight for the sources in our sample.

Although these sources are characterized by apparently low luminosities,
most of them are Compton thick, i.e.
their intrinsic luminosity must be much higher than observed
(by perhaps two orders of magnitude).
Yet, NGC 4941 is Compton thin and its intrinsic luminosity (obtained
by correcting for the estimated N$_H$) is only 
$\sim 2\times 10^{41}$ erg/sec, similar to some LINER nuclei.

It should be noted that
about half of the Seyferts in the Maiolino and Rieke's (\cite{maiolino_a})
sample have
[OIII] luminosities lower than NGC 4941, and have not been studied
in the X~rays. Therefore, the hard X--ray spectral properties of the 
lowest luminosity AGN population have still to be probed.

\subsection{The distribution of absorbing column densities} \label{disc_Nh}

The most remarkable result of our survey is that all the objects
are heavily obscured with N$_H > 10^{23.6}$cm$^{-2}$ and,
in particular, 6 out of 8 objects are Compton thick. More specifically 
NGC2273, NGC3393, NGC4939, NGC5643 and MCG-05-18-002 
were identified as Compton thick with N$_H > 10^{25}$cm$^{-2}$,
NGC1386 is Compton thick with N$_H > 10^{24}$cm$^{-2}$,
while NGC3081 and NGC 4941
were identified as Compton thin with N$_H \sim 5\times 10^{23}$cm$^{-2}$.
However, the nature of NGC4939, NGC4941 and MGC-05-18-002
is still questionable.

This result has to be compared with former spectral surveys. 
Smith \& Done (\cite{smith}) studied the spectra of
a sample of type 2 and 1.9 Seyferts observed with Ginga,
that were probably selected amongst bright X--ray sources.
The distribution of N$_H$ in their sample is shown in Fig.~\ref{fig_nh}.
The average absorbing column density is about $10^{22.6}$cm$^{-2}$.
Turner et al. (\cite{turner_a},\cite{turner_b}) analyzed  the spectra
of a sample of Sy2s from the ASCA archive. The latter sample includes a
larger fraction of
weak Sy2s with respect to the Ginga survey (see discussion in the
former section). As a consequence, it contains a larger 
fraction of heavily absorbed Sy2s, as shown in Fig.~\ref{fig_nh}.
\footnote{In this Figure we consider NGC4945 and Mk3 as having
$10^{24}<$N$_H<10^{25}$cm$^{-2}$ (Done et al. \cite{done}, Turner et al.
\cite{turner_b}),
NGC1068 with
N$_H>10^{25}$cm$^{-2}$ (Matt \cite{matt_b}, Matt et al. \cite{matt_d}),
Mk463E and NGC6240 with N$_H>10^{24}$cm$^{-2}$,
though the
column density absorbing the latter two objects is still matter of debate
(a more detailed discussion is given in Bassani et al. in prep.).}

Our sample was selected by means of an isotropic indicator
of the nuclear luminosity, the dereddened [OIII] flux,
and therefore is not biased against obscuration on the pc scale.
Thus, it contains a larger portion of heavily obscured Sy2s.
In particular, our survey doubles
the number of known Compton thick Seyferts:
before this study only
6 sources were surely known to be Compton thick, namely
NGC 1068, Circinus, NGC4945, NGC 6240,
NGC 6552, NGC7674 (the latter
identified by means of BeppoSAX as well, Ma\-la\-gu\-ti et al. \cite{malaguti}).

\begin{figure}[!]
\resizebox{\hsize}{!}{\includegraphics{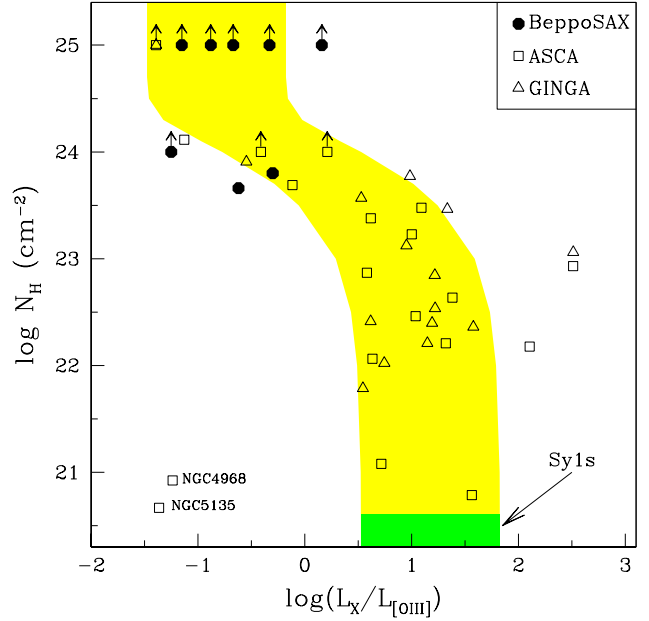}} 
 \caption{Distribution of the absorbing column density N$_H$
as a function of ratio between
the observed 2--10 keV luminosity and the (reddening corrected) [OIII]
luminosity. Filled circles are Sy2s observed by BeppoSAX presented in this
paper, open squares are Sy2s observed by ASCA
(Turner et al. \cite{turner_a}) and open triangles are Sy2s observed
by GINGA (Smith \& Done \cite{smith}). The shaded region indicates the
expected correlation by assuming that L$_X$ (2--10 keV) is
absorbed by the N$_H$ reported on the Y-axis, starting from the average
L$_X$/L$_{[OIII]}$ ratio observed in Sy1s (the width of the shaded region
reflects the $\pm 1\sigma$ dispersion
 of the L$_X$/L$_{[OIII]}$ distribution in Sy1s)
and by assuming a 1\% reflected component.}
\label{nh_xoiii}
\end{figure}

It is interesting to compare the properties of our sources with
respect to former surveys also in terms of their location on
the N$_H$ vs. L$_X$/L$_{[OIII]}$ diagram, as shown in 
Fig.\ref{nh_xoiii}\footnote{In Fig.\ref{nh_xoiii} we have excluded NGC2992,
NGC1667 and NGC1808, for long term variability affects the
L$_X$/L$_{[OIII]}$ ratio in these objects, as discussed in Bassani et al.
(in prep.). The values of L$_{[OIII]}$ are taken from Dadina et al.
(\cite{dadina}).}. As expected (see discussion in Sect.\ref{other_abs}),
N$_H$ correlates with the L$_X$/L$_{[OIII]}$ ratio. The shaded region
indicates the expected correlation by assuming that L$_X$ (2--10 keV) is
absorbed by the N$_H$ reported on the Y-axis, starting from the average
L$_X$/L$_{[OIII]}$ ratio observed in Sy1s (the width of the shaded region
reflects the $\pm 1\sigma$ dispersion
 of the L$_X$/L$_{[OIII]}$ distribution in Sy1s)
and by assuming a 1\% reflected component.
The observed distribution is in agreement with what expected from the
simple model outlined above, especially if allowance is made for an
efficiency of the reflected component different from what assumed.
Two objects, namely NGC4968 and NGC5135, are clearly out of the
correlation;
however, as discussed in detail in Bassani et al. (in prep.), these two
sources are very likely Compton thick
with N$_H >$10$^{24}$cm$^{-2}$. All of the Sy2s in our sample (filled circles)
lie in the high--N$_H$ and low--L$_X$/L$_{[OIII]}$ region of the
distribution, while most of the Sy2s
in former ASCA (open squares, Turner et al.
\cite{turner_a}) and GINGA (open triangles, Smith \& Done \cite{smith}) surveys
are in the low--N$_H$ and high--L$_X$/L$_{[OIII]}$ region.

One of the consequences of our result is that the average absorbing column
density in Sy2s turns out to be higher than what was estimated in the past.
However, the N$_H$ distribution in our sample alone
does not necessary reflect the real distribution.
Indeed, our sample is not biased 
against heavily absorbed nuclei, but is probably biased against
little absorbed Sy2s, since we avoided Sy2s already observed in 
previous surveys (i.e. generally X--ray bright sources). The issue of the real 
distribution of N$_H$ is tackled in Bassani et al. (in prep.), where the
spectra of several Sy2s (including the ones in this paper)
are collected and analyzed.

\subsection{Warm scattering versus cold reflection}
In Compton thick Seyferts the nuclear continuum can be observed only when
scattered by a warm, highly ionized mirror, or Compton reflected
by a cold neutral medium (possibly the molecular torus).
Since our survey has significantly enlarged the number of known Compton thick
Sy2s, it is now possible to statistically assess the relative importance
of the cold versus warm scattering in this class of objects.

By merging our sample with Compton thick sources previously reported in
the literature, we found that most of the 
Compton thick Seyferts are cold reflection
dominated. More specifically, out of 12 Compton thick Seyferts 7 are cold
reflection dominated (namely NGC 1386, NGC 2273, NGC 3393, NGC 4939, Circinus,
NGC 6552, NGC 7674), in 3 sources warm scattered and
Compton reflected components contribute to a similar extent (namely NGC 1068,
NGC 6240, and NGC4945), and
only one source (NGC 5643, though to be confirmed)
appears to be warm scattering dominated; the nature of the
scatterer in MGC-05-18-002 is still uncertain.

The apparent overabundance of cold
reflection dominated Compton thick sources must
probably be ascribed to the low efficiency of the free electron
scattering process. Indeed, in the latter case the fraction of scattered
light is given by

\begin{equation} \label{eq1}
f_{scatt} = N_e \sigma _T \Omega /4\pi
\end{equation}
where $N_e$ is the free electron column density and $\Omega$
is the solid angle subtended by the warm mirror. The average opening angle of
the light cones in Sy2s is $\Omega \approx 0.2$ (Maiolino \& Rieke
\cite{maiolino_a}).
If we assume that the warm mirror in Sy2s is the analogous of the
warm absorber observed in Sy1s, then $N_e\approx 10^{20}-10^{21}$cm$^{-2}$
(Reynolds \cite{reynolds}).
According to Eq.~\ref{eq1} this implies that the scattering
efficiency of the warm mirror is only $f_{scatt}\approx 10^{-4}$, i.e.
significantly lower than the typical efficiencies observed 
in Compton thick sources. The Compton cold reflection efficiency can be 
as high as 16\% in the 2--10 keV range, though
geometry dependent absorption effects (e.g. torus self shielding) can
lower this value to a few percent.

Those (few) Compton thick objects whose spectrum shows evidence for a warm
scattered component could either be characterized by an anomalously high
column density of the warm mirror, or the Compton reflected component
could be significantly absorbed along our line of sight; the latter
case might occur if the putative torus is edge--on.

\subsection{The soft excess} \label{disc_soft}

Most of the objects in our sample show an emission below $\sim 3$ keV
in excess of the extrapolation of the high energy spectrum.
Such soft X--ray excesses are commonly observed in type 2 Seyferts and 
have been studied by several authors.
This component is
usually not observed in type 1 Seyferts, presumably because
the low absorption affecting them makes
their flat power law component overwhelming.
The nature of the soft excess in type 2 Seyferts is not clear yet.
There are two possible origins: 1) thermal emission from hot
gas in the Narrow Line Region,
or 2) emission associated to starburst
activity in the host galaxy (X--ray binaries, supernova remnants, starburst
driven superwinds). The possibility that the observed soft excess is
due to the primary AGN
radiation scattered by a warm medium is not favored based on
energy budget arguments (Wilson \& Elvis \cite{wilson}).

Our data are not well suited to study the emission below 3 keV, since
at low energies other satellites (ROSAT, ASCA) are much more sensitive.
However, the wide spectral coverage of our BeppoSAX data enables us to 
determine what fraction of the soft X--ray emission is contributed by the flat
AGN component extrapolated to low energies, and what fraction, instead,
is actually due to an extra component. In other words, by means of our
fits in Sect.~\ref{results}
we can derive the flux of the black body component alone.

\begin{figure}[!]
\resizebox{\hsize}{!}{\includegraphics{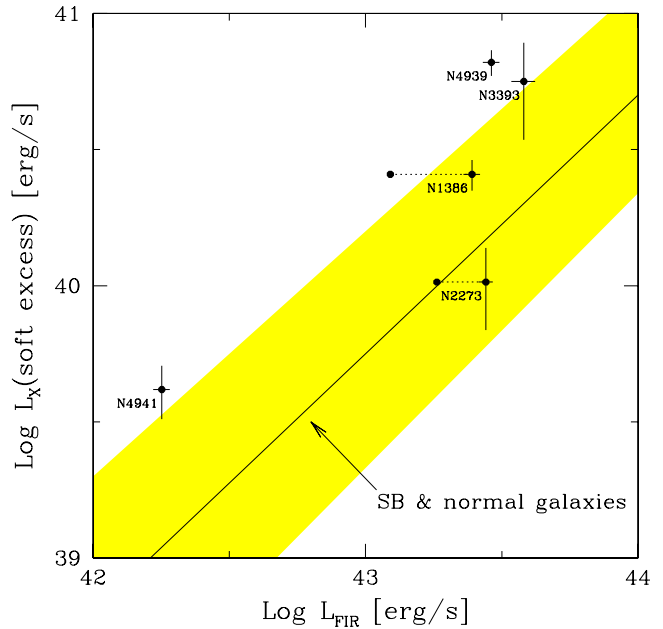}} 
 \caption{Distribution of the 0.5--4.5 keV excess emission versus the Far--IR
luminosity for the objects in our sample. The points connected by
dotted lines indicate lower limits for the starburst contribution to the
FIR luminosity (see text). The thick solid line indicates the
L$_{(0.5-4.5keV)}$--L$_{FIR}$ relationship of starburst and normal
galaxies; the shaded region indicates the 90\% confidence limits.}
\label{fig_softx}
\end{figure}

In these low luminosity Seyfert nuclei most of the Far--IR luminosity can
be ascribed to star formation in the host galaxy (star forming activity is
much more effective than the AGN in powering the FIR emission). So,
the FIR luminosity can be used to estimate the level of star formation,
and to determine the contribution of the latter to the observed soft excess. 
David et al. (\cite{david}) derived an empirical relationship
between L$_{0.5-4.5keV}$ and L$_{FIR}$ in starburst and normal galaxies.
This relation is shown in Fig.~\ref{fig_softx} by means of a thick solid line,
while the shaded region
indicates the 90\% confidence area. The same Figure shows the location of
the objects in our sample (NGC3081 and MGC-05-18-002 do not have available
IRAS data, while NGC5643 does not show evidence of a soft X--ray excess).
NGC 2273 seems to follow the relation for SB and normal galaxies, thus in this
object most of the excess can be completely ascribed to the starburst.
Most of the other objects show indications of 0.5--4.5 keV emission in 
excess of what would be expected from the starburst activity
alone, although NGC3393, NGC 1386 and NGC4941 are also consistent with the upper
limit of the SB/normal galaxies' distribution, when allowance is made for
uncertainties. NGC4939 (which also has the highest signal--to--noise spectrum)
is the object for which is strongest the evidence for an extra
contribution to the observed soft X--ray excess, i.e. emission related to the
AGN.

We should mention that, although weak, these Seyfert nuclei might contribute
some of the FIR luminosity. In the latter case the observed FIR luminosity
provides an upper limit to the contribution from the starburst.
Maiolino et al. (\cite{maiolino_b}) addressed the issue of the relative
importance of the IR emission from AGNs and from star
forming activity in their host galaxies.
Two of the objects in our sample, namely NGC 2273 and NGC 1386, were also
studied in Maiolino et al. (\cite{maiolino_b}).
By using their results we could
provide lower limits for the starburst contribution to the
FIR luminosity in these two objects; they are indicated by
points connected by dotted lines in Fig.~\ref{fig_softx}.
In the case of NGC 2273 the lower limit indicates that the soft X--ray excess
is fully consistent with it being originated by the starburst activity.
In the case of NGC 1386 it provides further support to the idea that
a fraction of the soft excess comes from the AGN (NLR) component.

\section{Conclusions}

We presented X--ray observations in the 0.1--100 keV spectral band
of 8 Seyfert 2 galaxies. The sources were selected from the Maiolino \&
Rieke (\cite{maiolino_a})
Seyfert sample (limited in B magnitude of the host galaxy)
according to their [OIII] flux, that should deliver a sample not biased
against heavily absorbed nuclei on the pc scale. Also, we avoided sources
already observed in former hard X--ray surveys, generally X--ray bright.
As a result, most of the sources in our sample are relatively X--ray weak.

We detected all of our targets in the 2--10 keV range, and two of them also 
in the 20--100 keV range.
Most of these X--ray weak AGNs are characterized
by prominent iron lines (at 6.4--7 keV) and
flat {\it observed} continua, indicating that they are powered by
the same kind of engine that powers X--ray bright
AGNs. However, these X--ray weak AGNs are characterized by larger
EW(Fe K$\alpha$) and flatter {\it observed} continua.

All of these sources are absorbed by column densities larger than
$4\times  10^{23}$ cm$^{-2}$ along our line of sight. Most of them appear
to be thick to Compton scattering with N$_H > 10^{25}$ cm$^{-2}$. In the
latter cases the observed continuum can generally be ascribed to Compton
reflection by cold material.

These findings support the unified model, and point to an
average absorbing column density in Sy2s that is
much higher than presumed on the basis of previous
X--ray surveys. The latter result has important implications for
the synthesis of the X--ray background.

Finally, in some of these sources there is indication that the soft 
X--ray
excess cannot be fully ascribed to starburst activity in the host galaxy, and
a significant fraction of it is probably related to the active nucleus itself.

\begin{acknowledgements}
We thank the referees I. Lehmann and G. Hasinger for helpful comments.
This research made use of SAXDAS linearized and cleaned event
files produced at the BeppoSAX Science Data Center.\hfill\break
This work was partially supported by the Italian National Research Council (CNR)
and by the Italian Space Agency (ASI).
In particular MS, RM, and GR acknowledge the partial financial support
from ASI through grant ARS--96--66, while GZ acknowledges financial
support from ASI through grants 95--RS--152 and ARS--96--70.
\end{acknowledgements}

\begin{table*}[!]
\caption[]{Spectral fit by using a transmission model where $\Gamma$ is
a free parameter} \label{tab_apx}
\begin{tabular}{lccccccccc}
\hline
\hline
\vspace{.01truecm}\\
Source & $\Gamma$ & N$_H$ & \multicolumn{3}{c}{Fe Line} &
F$_{2-10keV}$ & $\chi ^2$/d.o.f.\\
   &   & & Energy$^b$ & Normaliz. & EW & & \\
    &   &  (10$^{23}$cm$^{-2}$) &
         (keV) & (10$^{-5}$ph/cm$^2$/s) & (keV) &
    (10$^{-13}$erg/cm$^2$/s) & \\
\hline
NGC 2273 & $0.7^{+0.5}_{-0.5}$ & $<0.3$ & $6.4^a$ & $2.8^{+0.7}_{-0.7}$
 & $3.8^{+1.1}_{-1.1}$ &  9.49  & 34.2/38 \\
 &  &  & $6.96^a$ & $1.0^{+0.5}_{-0.5}$ & $1.45^{+0.95}_{-0.56}$ &  & \\
NGC 3081 & $1.7^{+0.26}_{-0.35}$ & $6.6^{+1.8}_{-3.5}$ &
  $6.48^{+0.21}_{-0.19}$ & $1.4^{+0.9}_{-0.9}$
 & $0.61^{+0.39}_{-0.21}$ &  13.2 &  40.7/38 \\
NGC 3393 & $-0.35^{+0.50}_{-0.24}$ & $<7.0$ &
  $6.51^{+0.28}_{-0.18}$ & $1.13^{-0.65}_{+0.63}$ & $3.5^{+2.}_{-2.}$
 &  3.83 &  15.2/19 \\
NGC 4939 & $0.71^{+0.27}_{-0.17}$ & $1.5^{+0.8}_{-0.6}$ &
  $6.4^a$ & $1.19^{+0.66}_{-0.54}$
 & $0.61^{+0.23}_{-0.34}$ & 15.3 &  85.4/92 \\
 & & & $6.96^a$ & $1.25^{+0.53}_{-0.64}$
 & $0.64^{+0.27}_{-0.32}$ &  &  \\
NGC 5643 & $1.53^{+0.23}_{-0.26}$ & $<0.03$ &
  $6.46^{+0.19}_{-0.13}$ & $2.16^{+0.8}_{-1.1}$ & $1.8^{+0.8}_{-0.9}$
  & 13.3 &  30.8/32 \\
MCG-05-18-002 & $0.57^{+0.88}_{-0.92}$ & -- &
  -- & -- & -- & 7.7 &  12.9/9 \\
\hline
\hline
\end{tabular}
Notes: \vspace{-.2truecm}
\begin{list}{}{}
\item $^a$ frozen parameter;
\item $^b$ rest frame.
\end{list}
\end{table*}

\appendix

\section{Spectral fit models with a thawed photon index}

In Sect. \ref{results}
we froze the photon index to 1.7 to reduce the number of
free parameters; this was required by the low statistics of the continuum
data points in most of the spectra. However, many authors often fit hard
X--ray data with transmission models whose photon index is left free.
To allow for a fair comparison of our data with these studies, 
here we report the results obtained by fitting such transmission
models to our data (Table~\ref{tab_apx}). The only exceptions are NGC1386 and
NGC 4941, whose N$_H$ and $\Gamma$ turn out to be completely indetermined
once $\Gamma$ is thawed.

It is worth noting that most of the sources are characterized by a very flat
photon index ($\Gamma \le 0.7$); this result supports our interpretation that
the sources are Compton thick, dominated by a cold reflection component.
The only exceptions are NGC3081, that we indeed identified as Compton thin,
and NGC5643, that is thought to be Compton thick, warm scattering dominated.

\end{document}